\documentclass[12pt]{article}

\usepackage{scicite,epsfig,amssymb,amsmath,lscape}

\usepackage{times}
\usepackage{caption}

\newcommand{\swift}{{\it Swift}}

\newcommand{\mjd}{${\rm MJD}$}

\newdimen\sa  \newdimen\sb
\def\parcs{\sa=.07em \sb=.03em
     \ifmmode $\rlap{.}$^{\scriptscriptstyle\prime\kern -\sb\prime}$\kern -\sa$
     \else \rlap{.}$^{\scriptscriptstyle\prime\kern -\sb\prime}$\kern -\sa\fi}

\newenvironment{sciabstract}{%
\begin{quote} \bf}
{\end{quote}}

\newcounter{lastnote}

\title{ASASSN-15lh: \\A Highly Super-Luminous Supernova} 

\author
{Subo Dong,$^{1\ast}$
B.~J.~Shappee,$^{2,3}$
J.~L.~Prieto,$^{4,5}$
S.~W.~Jha,$^{6}$
K.~Z.~Stanek,$^{7,8}$\\
T.~W.-S.~Holoien,$^{7,8}$ 
C.~S.~Kochanek,$^{7,8}$
T.~A.~Thompson,$^{7,8}$
N. Morrell,$^{9}$\\
I. B. Thompson,$^{2}$
U.~Basu,$^{7}$
J.~F.~Beacom,$^{7,8,10}$
D.~Bersier, $^{11}$
J.~Brimacombe,$^{12}$\\
J.~S.~Brown,$^{7}$
F.~Bufano,$^{13}$
Ping Chen,$^{14}$
E.~Conseil,$^{15}$
A.~B.~Danilet,$^{7}$
E.~Falco,$^{16}$\\
D.~Grupe,$^{17}$
S.~Kiyota,$^{18}$
G.~Masi,$^{19}$
B.~Nicholls,$^{20}$
F.~Olivares E.,$^{21,5}$
G.~Pignata,$^{21,5}$\\
G.~Pojmanski,$^{22}$
G.~V.~Simonian,$^{7}$
D.~M.~Szczygiel,$^{22}$ 
P.~R.~Wo\'zniak$^{23}$
\\}

\date{}

\begin{document} 


\baselineskip24pt


\maketitle 

\noindent
\normalsize{$^{1}$ Kavli Institute for Astronomy and Astrophysics, Peking University, Yi He Yuan Road 5, Hai Dian District, Beijing 100871, China}\\ 
\normalsize{$^{2}$ Carnegie Observatories, 813 Santa Barbara Street, Pasadena, CA 91101, USA}\\
\normalsize{$^{3}$ Hubble and Carnegie-Princeton Fellow}\\
\normalsize{$^{4}$ N\'ucleo de Astronom\'ia de la Facultad de Ingenier\'ia, Universidad Diego Portales, Av. Ej\'ercito 441, Santiago, Chile} \\
\normalsize{$^{5}$ Millennium Institute of Astrophysics, Santiago, Chile} \\
\normalsize{$^{6}$ Department of Physics and Astronomy, Rutgers, The State University of New Jersey, 136 Frelinghuysen Road, Piscataway, NJ 08854, USA}\\
\normalsize{$^{7}$ Department of Astronomy, The Ohio State University, 140 W.\ 18th Ave., Columbus, OH 43210, USA}\\
\normalsize{$^{8}$ Center for Cosmology and AstroParticle Physics (CCAPP), The Ohio State University, 191 W.\ Woodruff Ave., Columbus, OH 43210, USA }\\
\normalsize{$^{9}$ Las Campanas Observatory, Carnegie Observatories, Casilla 601, La Serena, Chile,}\\
\normalsize{$^{10}$ Department of Physics, The Ohio State University, 191 W. Woodruff Ave., Columbus, OH 43210, USA} \\
\normalsize{$^{11}$ Astrophysics Research Institute, Liverpool John Moores University, 146 Brownlow Hill, Liverpool L3 5RF, UK }\\
\normalsize{$^{12}$ Coral Towers Observatory, Cairns, Queensland 4870, Australia}\\
\normalsize{$^{13}$ INAF-Osservatorio Astrofisico di Catania, Via S.Sofia 78, 95123, Catania, Italy}\\
\normalsize{$^{14}$ Department of Astronomy, Peking University, Yi He Yuan Road 5, Hai Dian District, 100871, P. R. China}\\
\normalsize{$^{15}$ Association Francaise des Observateurs d'Etoiles Variables (AFOEV), Observatoire de Strasbourg 11, rue de l'Université, F-67000 Strasbourg, France}\\
\normalsize{$^{16}$ Harvard-Smithsonian Center for Astrophysics, 60 Garden St., Cambridge, MA 02138, USA.}\\
\normalsize{$^{17}$ Department of Earth and Space Science, Morehead State University, 235 Martindale Dr., Morehead, KY 40351, USA}\\
\normalsize{$^{18}$ Variable Star Observers League in Japan (VSOLJ), 7-1 Kitahatsutomi, Kamagaya, Chiba 273-0126, Japan}\\
\normalsize{$^{19}$ The Virtual Telescope Project, Via Madonna de Loco 47, 03023 Ceccano, Italy}\\
\normalsize{$^{20}$ Mt Vernon Observatory, 6 Mt~Vernon pl,  Nelson,  New Zealand}\\
\normalsize{$^{21}$ Departamento Ciencias Fisicas, Universidad Andres Bello, Av. Republica 252, Santiago, Chile}\\
\normalsize{$^{22}$ Warsaw University Astronomical Observatory, Al. Ujazdowskie 4, 00-478 Warsaw, Poland}\\
\normalsize{$^{23}$ Los Alamos National Laboratory, Mail Stop B244, Los Alamos, NM 87545, USA}\\
\\
\normalsize{$^\ast$To whom correspondence should be addressed; E-mail:  dongsubo@pku.edu.cn.}

\baselineskip24pt
\begin{sciabstract}

We report the discovery of ASASSN-15lh (SN 2015L), 
which we interpret as the most luminous supernova yet found. At redshift $z =
0.2326$, ASASSN-15lh reached an absolute magnitude of $M_{u,{\rm AB}}
= -23.5\pm0.1$ and bolometric luminosity $L_{\rm bol} = (2.2\pm0.2)\times
10^{45}\,\,{\rm ergs \,\,\,s^{-1}}$, which is more than twice as luminous as
any previously known supernova. It has several major features
characteristic of the hydrogen-poor super-luminous
supernovae (SLSNe-I), whose energy sources and progenitors are
currently poorly understood.
In contrast to most previously known SLSNe-I that reside in star-forming dwarf galaxies, ASASSN-15lh appears to be hosted by a luminous galaxy ($M_K \simeq -25.5$) with little star formation.
In the 4 months since first detection, ASASSN-15lh radiated
$(1.1\pm0.2)\times10^{52}$\,ergs, challenging the magnetar model
for its engine.

\end{sciabstract}

Only within the past two decades has the most luminous class of
supernovae (super-luminous supernovae, SLSNe) been identified (e.g.,
\cite{galyam}).  Compared with the most commonly discovered SNe (Type
Ia), SLSNe are more luminous by over two magnitudes at peak and rarer
by at least 3 orders of magnitude \cite{rate}. Like normal SNe, SLSNe
are classified by their spectra as either SLSN-I (hydrogen-poor) or
SLSN-II (hydrogen-rich). Yet the physical characteristics of SLSNe may
not be simple extensions from their low-luminosity
counterparts\cite{galyam}. In particular, the power source for SLSNe-I
is poorly understood \cite{quimby11}.  Adding to the puzzle, SLSNe
tend to explode in low-luminosity, star-forming dwarf galaxies
\cite{neillhost,stollhost,lunnanhost}.  The recent advent of
wide-area, un-targeted transient surveys has made the systematic
discovery and investigation of the SLSNe population possible (e.g.,
see \cite{quimbyreview} and \cite{nicholl2015} and references therein).
 
The All-Sky Automated Survey for SuperNovae
(ASAS-SN\footnote{http://www.astronomy.ohio-state.edu/$\sim$assassin/};
\cite{shappee14}) scans the visible sky every 2--3 nights to depths of
$V \simeq16.5-17.3$\,mag using a global network of 14\,cm telescopes
\cite{shappee14} in an un-targeted search for new transients,
particularly bright supernovae.

On 2015 June 14 \footnote{UT dates are used throughout this paper},
ASAS-SN triggered on a new source located at $\textrm{RA}=22^{\rm
h}02^{\rm m}15.\!\!^{\rm{s}}45$
$\textrm{Dec}=-61^{\circ}39'34.\!\!''6$ (J2000), coinciding with a
galaxy of then unknown redshift, APMUKS(BJ) B215839.70$-$615403.9
\cite{maddox90}.  Upon confirmation with our follow-up telescopes, we
designated this new source ASASSN-15lh and published its coordinates
\cite{atel}.

By combining multiple epochs of ASAS-SN images we extended the detections to fainter
fluxes, finding pre-discovery images of ASASSN-15lh from May 8, 2015
($V= 17.39\pm0.23$ mag), and Figure~1 shows the light curve through 2015 September 19.  The ASAS-SN light curve peaked at $V = 16.9\pm0.1$ on
approximately $t_{\rm peak}\sim {\rm JD}2457179$ (June 05, 2015) based
on a parabolic fit to the lightcurve (dashed line, Fig. 1). Follow-up images were taken with
the Las Cumbres Observatory Global Telescope Network (LCOGT) 1-m
telescopes, and the $BV$ light-curves with the galaxy contribution
subtracted are also shown.

We obtained an optical spectrum ($3700\--9200$ \AA) of ASASSN-15lh on 2015 June 21 with the du Pont 100-inch telescope. The
steep spectral slope with relatively high blue flux motivated {\swift}
UVOT/XRT \cite{burrows05} target-of-opportunity observations starting
on June 24, 2015. The 6-band {\swift} light curve spanning from the UV
to the optical (1928\,{\AA}$-$5468\,{\AA}) is shown in Figure~1. The {\swift}
spectral energy distribution (SED), peaking in the UV, indicates the
source has a high temperature. We derive a $3\sigma$ X-ray flux limit
of $< 1.6\times10^{-14} \,\,{\rm ergs\,s^{-1}\,cm^{-2}}$ (0.3-10
keV) from a total XRT exposure of 81\,ks taken between 2015 June 24
and Sep 18.

The du Pont spectrum is mostly featureless (Figure~2, panel (A), first
from the top)
except for a deep, broad absorption trough near $\sim 5100$\,\AA\,(observer
frame). {\tt SNID} \cite{snid}, a commonly used SN classification software that
has a spectral library of most types of supernovae except SLSN, failed to find
a good SN match. However, we noticed a resemblance between the trough and a feature attributed to
O~{\tt \rm II} absorption near $4100$\,{\AA}\,(rest frame) in the
spectrum of PTF10cwr/SN2010gx, a SLSN-I at $z = 0.230$ \cite{quimby11,2010gx,diverseic}.  Assuming that the ASASSN-15lh absorption trough (FWHM $\sim10^4$\,km s$^{-1}$) was also due
to the same feature indicated a similar redshift of $z\sim0.23$.  An
optical spectrum ($3250\--6150$ \AA) obtained on the Southern African Large Telescope (SALT)
revealed a clear Mg~{\tt \rm II} absorption doublet ($\lambda\lambda$2796, 2803)
at $z=0.232$, confirming the redshift expected from our tentative line 
identification. Subsequent Magellan/Clay (July 6) and SALT (July 7)
spectra refined the redshift to $z = 0.2326$ (panels (C) and (D),
Figure~2). The available rest frame spectra show continua with steep
spectral slope, relatively high blue fluxes and several broad
absorption features also seen in PTF10cwr/SN2010gx (features `a', `b' and `c' labeled in panel (A) of Figure~2) and without hydrogen or helium
features, consistent with the main spectral features of SLSNe-I
\cite{quimby11,galyam}. The broad absorption feature near
$4400$\,{\AA}\, (`d' in Figure~2) seen in PTF10cwr/SN2010gx is not
present in ASASSN-15lh. ASASSN-15lh thus has some distinct spectral
characteristics in comparison with PTF10cwr/SN2010gx and some other SLSNe-I \cite{quimby11}.
 
Using a luminosity distance of $1171 {\,\rm Mpc}$ (standard {\it
Planck} cosmology at $z=0.2326$), Galactic extinction of $E(B-V) =
0.03$\,mag \cite{ext}, assuming no host extinction (thus the luminosity derived is likely a lower limit), and fitting the
{\swift} and LCOGT flux measurements to a simple blackbody (BB) model, we obtain
declining rest-frame temperatures of $T_{\rm BB}$ from $2.1
\times10^4$\,K to $1.3 \times 10^4$\, K and bolometric luminosities of
$L_{\rm bol} = 2.2\times 10^{45}$ to $0.4\times 10^{45} {\,\rm ergs\,\,s^{-1}}$ at
rest-frame phases relative to the peak of $t_{\rm rest} \sim 15$ and $\sim
50\,{\rm days}$, respectively (see Figure~3). ASASSN-15lh's bolometric magnitude declines
at a best-fit linear rate of $0.048\,{\rm mag/day}$, which is practically identical to   SLSN-I iPTF13ajg \cite{iptf13ajg} at $0.049\,{\rm mag/day}$ during similar phases $(\sim 10$ to $\sim
50\,{\rm days})$. Subsequently, the luminosity and temperature
reaches a ``plateau'' phase with slow changes and a similar trend is 
also seen for iPTF13ajg though with sparser coverage. Overall, the temperature and luminosity 
time evolution resemble
iPTF13ajg, but ASASSN-15lh has 
a systematically higher temperature at similar phases.  The estimated
BB radius of $\sim 5\times 10^{15} \,{\rm cm}$ near the peak is similar to those
derived for other SLSNe-I (see, e.g.,
\cite{quimby11,iptf13ajg}). These similarities in the evolution of
key properties support the argument that ASASSN-15lh is a member of the SLSN-I 
class but with extreme properties.  

The absolute magnitudes (AB) in the rest-frame $u$-band are shown in
Figure~4.  Using either $T_{\rm BB}$ or the spectra, there is little
K-correction \cite{kcorrection} in converting from $B$-band to rest-frame $u_{\rm AB}$
with ${\rm K}_{B\rightarrow u_{\rm AB}} = -0.1$.  The solid red points
at $t_{\rm rest} \gtrsim 10 \,{\rm days}$ include $B$-band data.  Before $\sim 10 {\,\rm days}$ we lack
measurements in blue bands. To estimate $M_{u,{\rm AB}}$ at these
earlier epochs we assume the $B-V=-0.3$\,mag color
and K-corrections found for the later epochs with {\swift}
photometry. We estimate an integrated bolometric luminosity radiated
of $\sim (1.1\pm0.2)\times10^{52} \,{\rm ergs\,}$ over $108 {\,\rm
days}$ in the rest frame.  Although our estimates at $t_{\rm
rest}\lesssim10$\,days should be treated with caution, we can securely
conclude that the peak $M_{u,{\rm AB}}$ is at or brighter than $-23.5\pm
0.1$, with a bolometric luminosity at or greater than $(2.2\pm0.2)\times
10^{45} {\,\rm ergs\,\,s^{-1}}$.  Both values are without precedent 
for any supernova recorded in the literature. In Figure~4, we compare ASASSN-15lh with a
sample of SLSNe-I \cite{quimby11,iptf13ajg}.  Although its spectra 
resemble the SLSNe-I subclass, ASASSN-15lh stands out from the
luminosity distribution of known SLSNe-I, whose luminosities are
narrowly distributed around $M\sim -21.7$ \cite{rate,candle}.  In
Table~S1, we list the peak luminosities of the five most luminous SNe
discovered to date, including both SLSN-I and SLSN-II. The spectral correspondence and similarities in temperature, luminosity and radius 
evolutions between ASASSN-15lh and some SLSNe-I leads to the conclusion
that ASASSN-15lh is the most luminous supernova yet discovered. 
Even though we find that SLSN-I is the most plausible classification of 
ASASSN-15lh, it is important to consider other 
interpretations given its distinct properties.
We discuss alternative physical interpretations of ASASSN-15lh in the Supplementary text, 
and given all the currently available data, we conclude that it is most likely a
supernova, albeit an extreme one.

The rate of events with similar luminosities to 
ASASSN-15lh is
uncertain. Based on a simple model of transient light curves in
ASAS-SN observations tuned to reproduce the  
magnitude distribution of ASAS-SN Type Ia supernovae (see Supplementary text), 
the discovery of one ASASSN-15lh-like event implies a
rate of $r \simeq 0.6$\,Gpc$^{-3}$ yr$^{-1}$ (90\% confidence: $0.21 < r < 2.8$). 
This is at least 3 times and can be as much as 100 times smaller
than the overall rate of SLSNe-I, $r \simeq 32$\,Gpc$^{-3}$ yr$^{-1}$ (90\%
confidence: $6 < r < 109$) from \cite{rate}, and suggests a steeply falling
luminosity function for such supernovae.

For a redshift of $z=0.2326$, the host galaxy of ASASSN-15lh has $M_K
\approx -25.5$, which is much more luminous than the Milky Way. We
estimate an effective radius for the galaxy of $2.4\pm0.3$\,kpc and a
stellar mass of $M_{\star} \approx 2\times10^{11}$~M$_{\odot}$. This
is in contrast to the host galaxies of other SLSNe, which tend
to have much lower $M_\star$ (e.g.,
\cite{neillhost,stollhost,lunnanhost}). However, given the currently available data,
we cannot rule out the possibility that the host is a dwarf satellite galaxy seen in projection.
The lack of narrow hydrogen and oxygen emission lines from 
the galaxy superimposed in the supernova spectra
implies little star formation ${\rm SFR} < 0.3$\,M$_{\odot}$ yr$^{-1}$
by applying the conversions in \cite{starformation}.  LCOGT
astrometry places ASASSN-15lh within $0\parcs2$ (750\,pc) of the
center of the nominal host. A detailed discussion of the host properties 
is provided in the supplementary text.

The power source for ASASSN-15lh is unknown. Traditional mechanisms invoked 
for normal SNe likely cannot explain SLSNe-I \cite{quimby11}.  The lack of hydrogen or 
helium suggests that shock
interactions with hydrogen-rich circumstellar material, invoked to
interpret some SLSNe, cannot explain SLSNe-I or ASASSN-15lh. SLSN-I
post-peak decline rates appear too fast to be explained by the
radioactive decay of $^{56}$Ni \cite{quimby11} -- the energy source
for Type Ia supernovae. Both the decline rate of the late-time light curve 
and the integral method \cite{katz13} will allow tests of
whether ASASSN-15lh is powered by $^{56}$Ni, and we estimate that
$\gtrsim 30$\,M$_\odot$ of $^{56}$Ni would be required to produce
ASASSN-15lh's peak luminosity.  
Another possibility is
that the spindown of a rapidly-rotating, highly-magnetic neutron star
(a magnetar) powers the extraordinary emission \cite{bodenheimer_ostriker,kasen_bildsten,woosley}. To
match the peak $L_{\rm bol}$ and timescale of ASASSN-15lh, the
light-curve models of \cite{kasen_bildsten} imply a magnetar spin
period and magnetic field strength of $P\simeq1$\,ms and
$B\simeq10^{14}$\,G, respectively, assuming that all of the spindown
power is thermalized in the stellar envelope. If efficient
thermalization continues, this model predicts an $L_{\rm bol}\propto
t^{-2}$ power-law at late times. The total observed energy radiated so
far ($1.1\pm0.2\times10^{52}$\,ergs) strains a magnetar
interpretation because, for $P\lesssim1$\,ms, gravitational wave
radiation should limit the total rotational energy available to
$E_{\rm rot}^{\rm max}\sim3\times10^{52}$\,ergs \cite{metzger} and the
total radiated energy to a third of $E_{\rm rot}^{\rm max}$, which is
$\sim10^{52}$\,ergs \cite{kasen_bildsten}.

The extreme luminosity of ASASSN-15lh opens up the possibility of observing 
such supernovae in the early universe. An event similar to ASASSN-15lh 
could be observed with the Hubble Space Telescope out to $z\sim6$, and with the James Webb Space Telescope out to
$z\gtrsim10$ \cite{candle}. A well-observed local counterpart will be critical in
making sense of future observations of the transient high-redshift
universe.

\bibliography{scibib.bib}

\begin{thebibliography}{10}

\bibitem{galyam}
A.~{Gal-Yam}, {\it \it Science\/} {\bf 337}, 927 (2012).

\bibitem{rate}
R.~M. {Quimby}, F.~{Yuan}, C.~{Akerlof}, J.~C. {Wheeler}, {\it \it Mon. Not. R.
  Astron. Soc.\/} {\bf 431}, 912 (2013).

\bibitem{quimby11}
R.~M. {Quimby}, {\it et~al.\/}, {\it \it Nature\/} {\bf 474}, 487 (2011).

\bibitem{neillhost}
J.~D. {Neill}, {\it et~al.\/}, {\it \it Astrophys. J.\/} {\bf 727}, 15 (2011).

\bibitem{stollhost}
R.~{Stoll}, {\it et~al.\/}, {\it \it Astrophys. J.\/} {\bf 730}, 34 (2011).

\bibitem{lunnanhost}
R.~{Lunnan}, {\it et~al.\/}, {\it \it Astrophys. J.\/} {\bf 787}, 138 (2014).

\bibitem{quimbyreview}
R.~M. {Quimby}, {\it IAU Symposium\/}, A.~{Ray}, R.~A. {McCray}, eds. (2014),
  vol. 296 of {\it IAU Symposium\/}, pp. 68--76.

\bibitem{nicholl2015}
M.~{Nicholl}, {\it et~al.\/}, {\it \mnras\/} {\bf 452}, 3869 (2015).

\bibitem{shappee14}
B.~J. {Shappee}, {\it et~al.\/}, {\it \it Astrophys. J.\/} {\bf 788}, 48
  (2014).

\bibitem{maddox90}
S.~J. {Maddox}, G.~{Efstathiou}, W.~J. {Sutherland}, J.~{Loveday}, {\it
  \mnras\/} {\bf 243}, 692 (1990).

\bibitem{atel}
B.~{Nicholls}, {\it et~al.\/}, {\it The Astronomer's Telegram\/} {\bf 7642}, 1
  (2015).

\bibitem{burrows05}
D.~N. {Burrows}, {\it et~al.\/}, {\it \it Space Sci. Rev.\/} {\bf 120}, 165
  (2005).

\bibitem{snid}
S.~{Blondin}, J.~L. {Tonry}, {\it \it Astrophys. J.\/} {\bf 666}, 1024 (2007).

\bibitem{2010gx}
A.~{Pastorello}, {\it et~al.\/}, {\it \it Astrophys. J. L.\/} {\bf 724}, L16
  (2010).

\bibitem{diverseic}
C.~{Inserra}, {\it et~al.\/}, {\it \apj\/} {\bf 770}, 128 (2013).

\bibitem{ext}
E.~F. {Schlafly}, D.~P. {Finkbeiner}, {\it \apj\/} {\bf 737}, 103 (2011).

\bibitem{iptf13ajg}
P.~M. {Vreeswijk}, {\it et~al.\/}, {\it \it Astrophys. J.\/} {\bf 797}, 24
  (2014).

\bibitem{kcorrection}
D.~W. {Hogg}, I.~K. {Baldry}, M.~R. {Blanton}, D.~J. {Eisenstein}, {\it ArXiv
  Astrophysics e-prints:astro-ph/0210394\/}  (2002).

\bibitem{candle}
C.~{Inserra}, S.~J. {Smartt}, {\it \it Astrophys. J.\/} {\bf 796}, 87 (2014).

\bibitem{starformation}
S.~{Savaglio}, K.~{Glazebrook}, D.~{Le Borgne}, {\it \it Astrophys. J.\/} {\bf
  691}, 182 (2009).

\bibitem{katz13}
B.~{Katz}, D.~{Kushnir}, S.~{Dong}, {\it ArXiv e-prints, arXiv:1301.6766\/}
  (2013).

\bibitem{bodenheimer_ostriker}
P.~{Bodenheimer}, J.~P. {Ostriker}, {\it \apj\/} {\bf 191}, 465 (1974).

\bibitem{kasen_bildsten}
D.~{Kasen}, L.~{Bildsten}, {\it \apj\/} {\bf 717}, 245 (2010).

\bibitem{woosley}
S.~E. {Woosley}, {\it \apjl\/} {\bf 719}, L204 (2010).

\bibitem{metzger}
B.~D. {Metzger}, D.~{Giannios}, T.~A. {Thompson}, N.~{Bucciantini},
  E.~{Quataert}, {\it \mnras\/} {\bf 413}, 2031 (2011).

\bibitem{wiserep}
O.~{Yaron}, A.~{Gal-Yam}, {\it \it Pub. Astron. Soc. Pac.\/} {\bf 124}, 668
  (2012).

\bibitem{alard00}
C.~{Alard}, {\it \it Astron. Astrophys.\/} {\bf 144}, 363 (2000).

\bibitem{poole08}
T.~S. {Poole}, {\it et~al.\/}, {\it \it Mon. Not. R. Astron. Soc.\/} {\bf 383},
  627 (2008).

\bibitem{breeveld10}
A.~A. {Breeveld}, {\it et~al.\/}, {\it \mnras\/} {\bf 406}, 1687 (2010).

\bibitem{brownatel}
P.~J. {Brown}, {\it The Astronomer's Telegram\/} {\bf 8086}, 1 (2015).

\bibitem{xrayatel}
R.~{Margutti}, {\it The Astronomer's Telegram\/} {\bf 8089}, 1 (2015).

\bibitem{lang10}
D.~{Lang}, D.~W. {Hogg}, K.~{Mierle}, M.~{Blanton}, S.~{Roweis}, {\it \it
  Astronomical Journal\/} {\bf 139}, 1782 (2010).

\bibitem{henden15}
A.~A. {Henden}, S.~{Levine}, D.~{Terrell}, D.~L. {Welch}, {\it American
  Astronomical Society Meeting Abstracts\/} (2015), vol. 225 of {\it American
  Astronomical Society Meeting Abstracts\/}, p. 336.16.

\bibitem{pysalt}
S.~M. {Crawford}, {\it et~al.\/}, {\it Society of Photo-Optical Instrumentation
  Engineers (SPIE) Conference Series\/} (2010), vol. 7737 of {\it Society of
  Photo-Optical Instrumentation Engineers (SPIE) Conference Series\/}, p.~25.

\bibitem{drake}
A.~J. {Drake}, {\it et~al.\/}, {\it \it Astrophys. J.\/} {\bf 696}, 870 (2009).

\bibitem{VHS}
R.~G. {McMahon}, {\it et~al.\/}, {\it The Messenger\/} {\bf 154}, 35 (2013).

\bibitem{sextractor}
E.~{Bertin}, S.~{Arnouts}, {\it Astronomy and Astrophysics, Supplement\/} {\bf
  117}, 393 (1996).

\bibitem{des}
P.~{Melchior}, A.~{Drlica-Wagner}, K.~{Bechtol}, E.~{Rykoff}, W.~H.~D.~E.
  {Survey}, {\it The Astronomer's Telegram\/} {\bf 7843}, 1 (2015).

\bibitem{kriek09}
M.~{Kriek}, {\it et~al.\/}, {\it \apj\/} {\bf 700}, 221 (2009).

\bibitem{bc03}
G.~{Bruzual}, S.~{Charlot}, {\it \mnras\/} {\bf 344}, 1000 (2003).

\bibitem{galfit}
C.~Y. {Peng}, L.~C. {Ho}, C.~D. {Impey}, H.-W. {Rix}, {\it \it Astronomical
  Journal\/} {\bf 124}, 266 (2002).

\bibitem{chen13}
T.-W. {Chen}, {\it et~al.\/}, {\it \it The Astrophysical Journal Letters\/}
  {\bf 763}, L28 (2013).

\bibitem{lunnan15}
R.~{Lunnan}, {\it et~al.\/}, {\it \apj\/} {\bf 804}, 90 (2015).

\bibitem{chen15}
T.-W. {Chen}, {\it et~al.\/}, {\it \mnras\/} {\bf 452}, 1567 (2015).

\bibitem{ptf12dam}
R.~M. {Quimby}, {\it et~al.\/}, {\it The Astronomer's Telegram\/} {\bf 4121}, 1
  (2012).

\bibitem{sn2003du}
V.~{Stanishev}, {\it et~al.\/}, {\it \it Astron. Astrophys.\/} {\bf 469}, 645
  (2007).

\bibitem{sn1998bw}
A.~{Clocchiatti}, N.~B. {Suntzeff}, R.~{Covarrubias}, P.~{Candia}, {\it \it
  Astronomical Journal\/} {\bf 141}, 163 (2011).

\bibitem{sn1999em}
M.~C. {Bersten}, M.~{Hamuy}, {\it \apj\/} {\bf 701}, 200 (2009).

\bibitem{sn1987a}
N.~B. {Suntzeff}, P.~{Bouchet}, {\it \it Astronomical Journal\/} {\bf 99}, 650
  (1990).

\bibitem{milkyway}
T.~C. {Licquia}, J.~A. {Newman}, J.~{Brinchmann}, {\it \apj\/} {\bf 809}, 96
  (2015).

\bibitem{csk06}
C.~S. {Kochanek}, {\it Saas-Fee Advanced Course 33: Gravitational Lensing:
  Strong, Weak and Micro\/}, G.~{Meylan}, {\it et~al.\/}, eds. (2006), pp.
  91--268.

\bibitem{MacLeod14}
C.~L. {MacLeod}, {\it et~al.\/}, {\it \apj\/} {\bf 721}, 1014 (2010).

\bibitem{vandenberk04}
D.~E. {Vanden Berk}, {\it et~al.\/}, {\it \apj\/} {\bf 601}, 692 (2004).

\bibitem{agnbook}
B.~M. {Peterson}, {\it {An Introduction to Active Galactic Nuclei}\/} (1997).

\bibitem{css10}
A.~J. {Drake}, {\it et~al.\/}, {\it \it Astrophys. J.\/} {\bf 735}, 106 (2011).

\bibitem{ptftde}
I.~{Arcavi}, {\it et~al.\/}, {\it \apj\/} {\bf 793}, 38 (2014).

\bibitem{pstde}
S.~{Gezari}, {\it et~al.\/}, {\it \it Nature\/} {\bf 485}, 217 (2012).

\bibitem{Strubbe2015}
L.~E. {Strubbe}, N.~{Murray}, {\it ArXiv e-prints:1509.04277\/}  (2015).

\bibitem{14ae}
T.~W.-S. {Holoien}, {\it et~al.\/}, {\it \mnras\/} {\bf 445}, 3263 (2014).

\bibitem{14li}
T.~W.-S. {Holoien}, {\it et~al.\/}, {\it ArXiv e-prints:1507.01598\/}  (2015).

\bibitem{dougie}
J.~{Vink{\'o}}, {\it et~al.\/}, {\it \apj\/} {\bf 798}, 12 (2015).

\bibitem{Chornock14}
R.~{Chornock}, {\it et~al.\/}, {\it \apj\/} {\bf 780}, 44 (2014).

\bibitem{haring}
N.~{H{\"a}ring}, H.-W. {Rix}, {\it \apjl\/} {\bf 604}, L89 (2004).

\bibitem{css12}
S.~{Benetti}, {\it et~al.\/}, {\it \it Mon. Not. R. Astron. Soc.\/} {\bf 441},
  289 (2014).

\bibitem{sn2008es}
A.~A. {Miller}, {\it et~al.\/}, {\it \it Astrophys. J.\/} {\bf 690}, 1303
  (2009).

\end{thebibliography}

\bibliographystyle{Science}

\section*{Acknowledgement}
We acknowledge: Bing Zhang, Luis Ho, Avishay Gal-Yam, Boaz Katz for
comments; NSF AST-1515927, OSU CCAPP, Mt.~Cuba
Astronomical Foundation, TAP,  SAO, CAS Grant No. XDB09000000 (SD);
NASA Hubble Fellowship (BJS); FONDECYT grant 1151445, MAS project
IC120009 (JLP); NSF CAREER award AST-0847157 (SWJ); DOE
DE-FG02-97ER25308 (TWSH); NSF PHY-1404311 (JFB); FONDECYT 3140326 (FOE) and LANL Laboratory
Directed Research and Development program (PRW).
All data used in this paper are made public including the photometric data (see Supplement tables) and spectroscopic data are available at public repository WISeREP \cite{wiserep} (http://wiserep.weizmann.ac.il). Materials and methods are available as supplementary materials on Science Online.


\clearpage

\includegraphics*[width=0.98\textwidth]{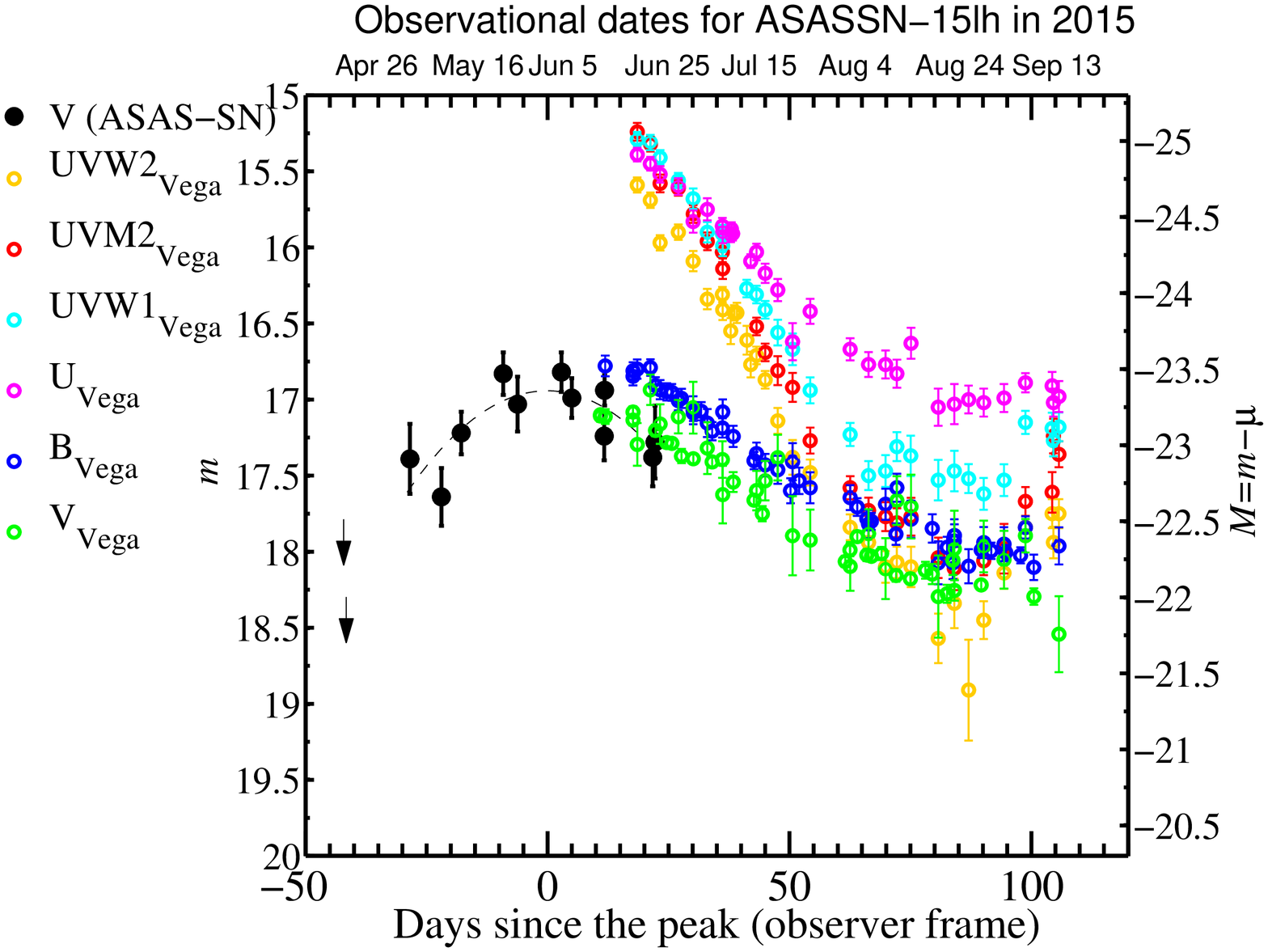}
\noindent{\bf Figure~1.}
{{\bf Multi-band light curve of ASASSN-15lh.}
The $V$-band ASAS-SN light curve is shown as black solid dots, 
and upper limits are represented by black arrows. 
{\swift} and LCOGT 1-m data are shown as open circles.
 
\clearpage
 
\includegraphics*[width=1.0\textwidth]{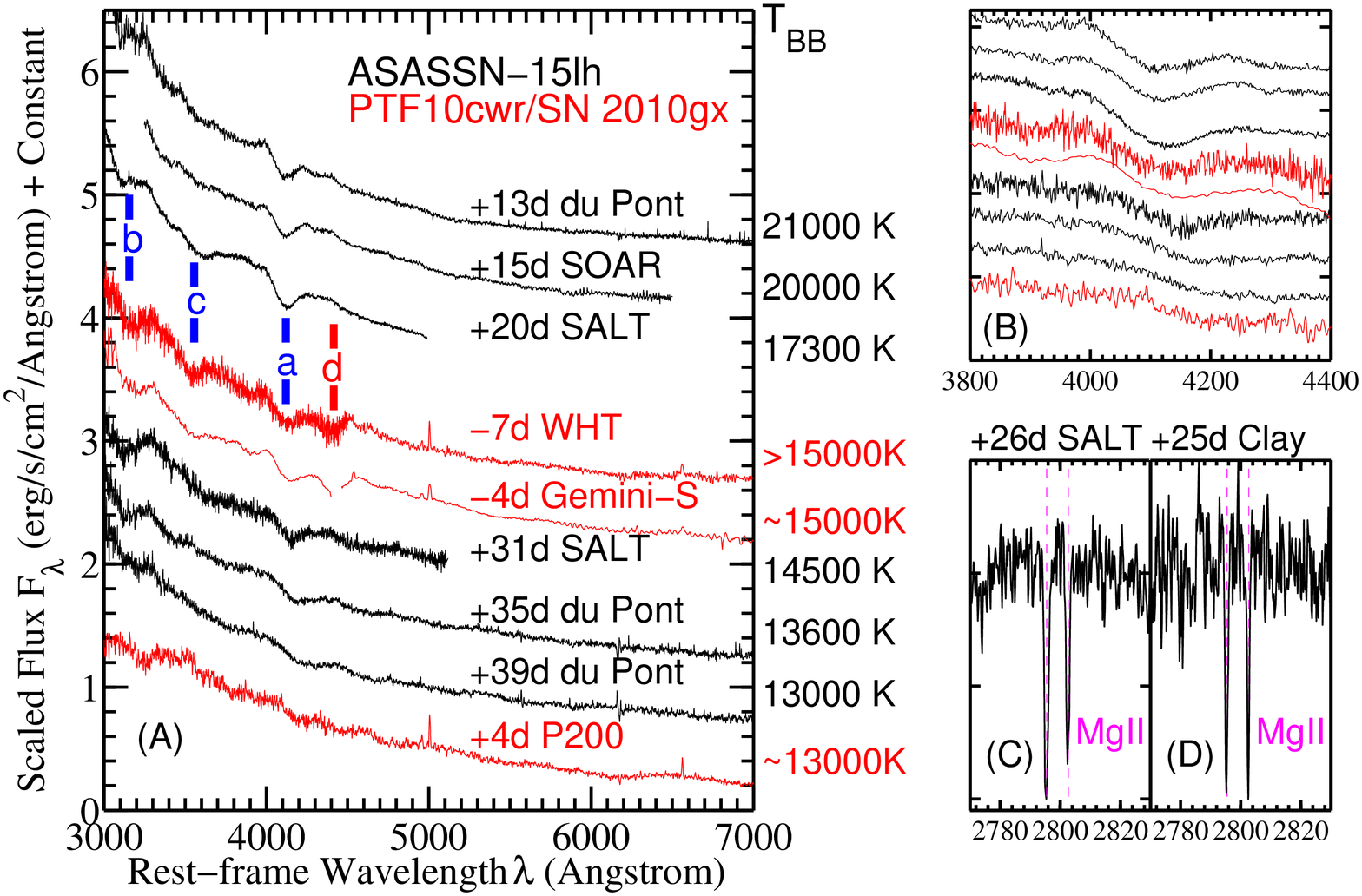}
\noindent{\bf Figure~2.}  {\bf Rest-frame spectra of ASASSN-15lh (black)
compared with SLSN-I PTF10cwr/SN 2010gx (red).} (A): The spectra
are offset for clarity,  
labelled with phases and telescopes, and they are ranked
by descending $T_{\rm BB}$ (given on the right) from the top.  The ASASSN-15lh spectra are
blue and featureless, except for broad absorption features labelled
`a', `b' and `c' (marked in blue), which match those of
PTF10cwr/SN 2010gx at similar $T_{\rm BB}$. Absorption features `a' at
$\sim4100$\,\AA\,and `d' at $\sim4400$\,\AA\, (marked in red) in
PTF10cwr/SN 2010gx are commonly attributed to O~{\tt \rm II} (e.g., \cite{quimby11,diverseic}). The
$\sim4400$\,\AA\,feature is not present in ASASSN-15lh.  Panel (B)
shows close-ups of the $4100$\,\AA\, features, whose evolution in
shape, depth and velocity as a function of $T_{\rm BB}$ is similar for
both supernovae. The ASASSN-15lh host redshift ($z=0.2326$) is
determined from the Mg~{\tt \rm II} doublets seen in the SALT and Clay MagE
spectra (panels (C) and (D)), with EW $ 0.55\pm0.05$\AA\, and
$0.49\pm0.05$\AA\,, respectively.
\clearpage

\includegraphics*[width=0.8\textwidth]{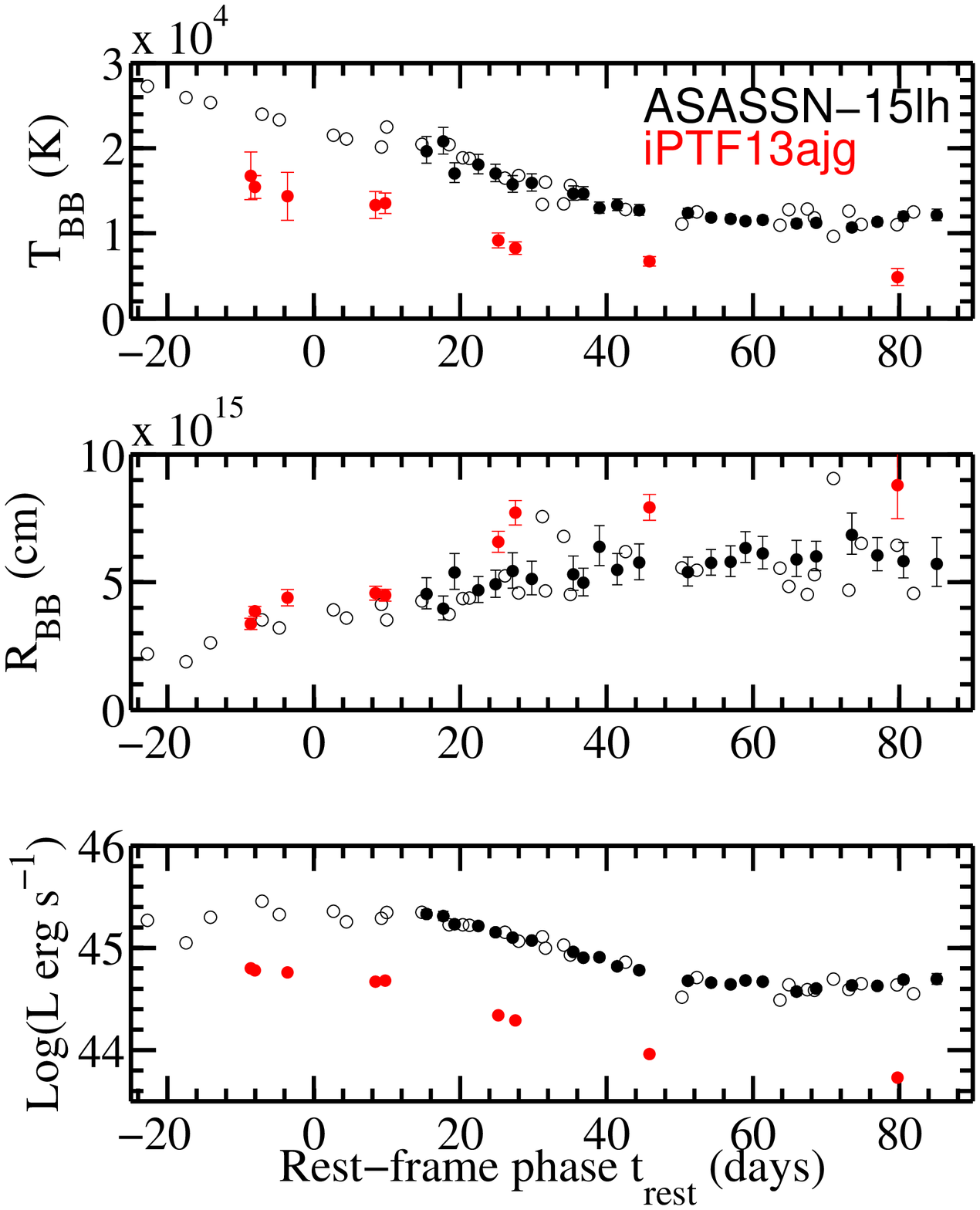}
\newline
\noindent{\bf Figure~3.}  {\bf Time evolution of blackbody temperatures,
radii and bolometric luminosities for ASASSN-15lh (black) and SLSN-I
iPTF13ajg (red).} Solid black dots show estimates derived from the full UV
and optical bands whereas the open circles show those from optical
only. For $t_{\rm rest}< 10\,{\rm days}$, only V-band is availablFe and
the temperatures are estimated based on linear extrapolation from $\mjd =
57191 - 57241$.
\clearpage

\includegraphics*[width=1.0\textwidth]{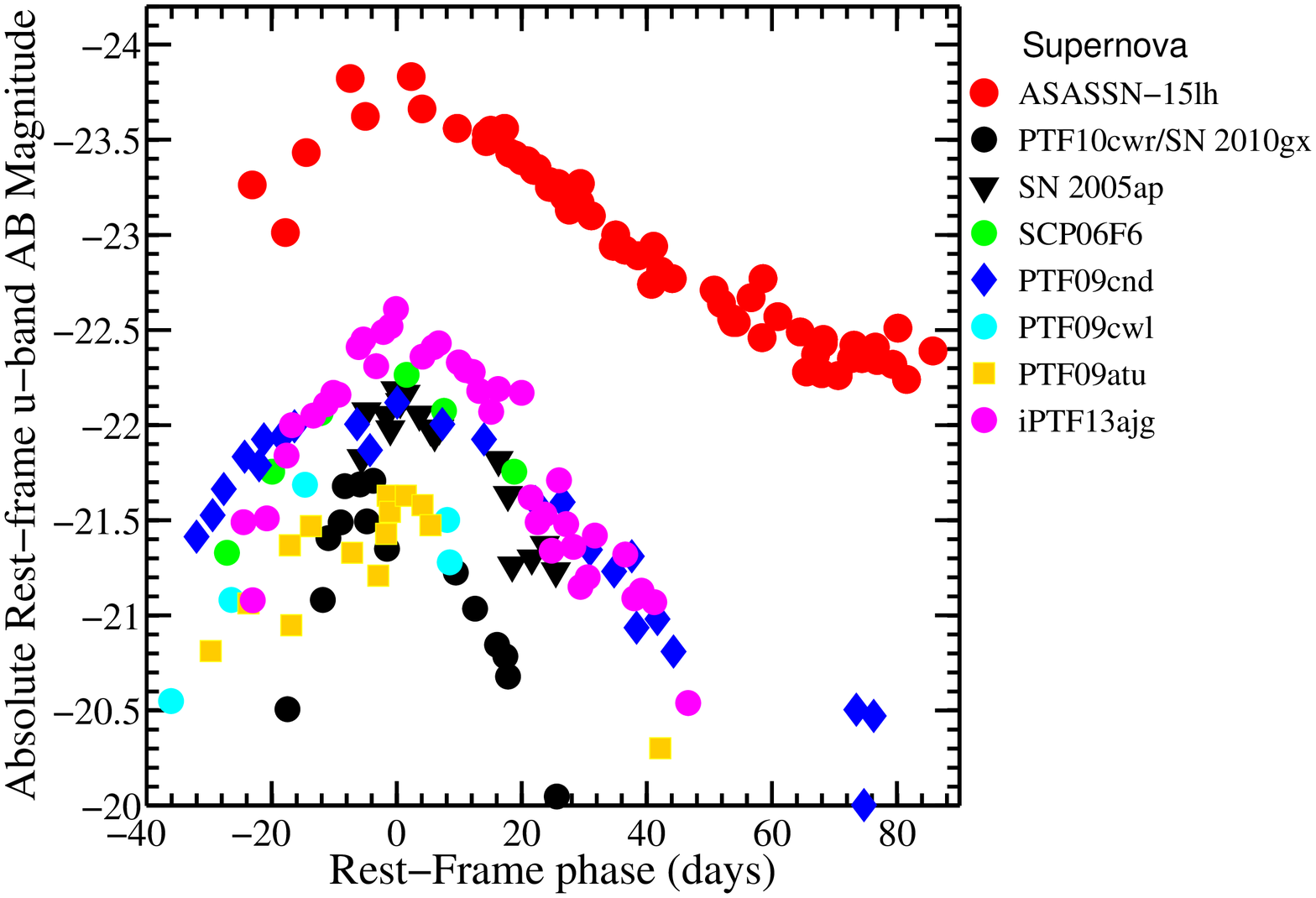}
\noindent{\bf Figure~4.}  {\bf Rest-frame absolute magnitude light curve of
ASASSN-15lh near peak compared with other SLSNe-I.} Estimates of
$M_{u,{\rm AB}}$ for ASASSN-15lh at $t_{\rm rest}\gtrsim 10\,{\rm days}$ 
are derived from $B$-band fluxes, which are subject to small
K-corrections, whereas the less reliable $M_{u,{\rm AB}}$
estimates are based on $V$-band only for $t_{\rm rest}\lesssim 10\,{\rm days}$. The comparison sample \cite{quimby11,iptf13ajg} includes the
most luminous SLSNe-I previously known. At $M_{u,{\rm AB}} = -23.5$,
ASASSN-15lh stands out from the SLSNe-I luminosity distribution
\cite{rate,candle}. Its peak bolometric absolute magnitude is more
than $\sim 1\,$mag more luminous than any other SLSN-I.

\clearpage
\setcounter{page}{1}
\begin{center}
\title{{\LARGE Supplementary Materials for}\\[0.5cm]
{\bf\large ASASSN-15lh: A Highly Super-Luminous Supernova}} 
\newline
\author
{Subo Dong,$^{1\ast}$
B.~J.~Shappee,$^{2,3}$
J.~L.~Prieto,$^{4,5}$
S.~W.~Jha,$^{6}$
K.~Z.~Stanek,$^{7,8}$\\
T.~W.-S.~Holoien,$^{7,8}$ 
C.~S.~Kochanek,$^{7,8}$
T.~A.~Thompson,$^{7,8}$
N. Morrell,$^{9}$\\
I. B. Thompson,$^{2}$
U.~Basu,$^{7}$
J.~F.~Beacom,$^{7,8,10}$
D.~Bersier, $^{11}$
J.~Brimacombe,$^{12}$\\
J.~S.~Brown,$^{7}$
F.~Bufano,$^{13}$
Ping Chen,$^{14}$
E.~Conseil,$^{15}$
A.~B.~Danilet,$^{7}$
E.~Falco,$^{16}$\\
D.~Grupe,$^{17}$
S.~Kiyota,$^{18}$
G.~Masi,$^{19}$
B.~Nicholls,$^{20}$
F.~Olivares E.,$^{21,5}$
G.~Pignata,$^{21,5}$\\
G.~Pojmanski,$^{22}$
G.~V.~Simonian,$^{7}$
D.~M.~Szczygiel,$^{22}$ 
P.~R.~Wo\'zniak$^{23}$
\newline
\\}
\normalsize{Correspondence to: dongsubo@pku.edu.cn.}
\newline
\end{center}

{{\bf This PDF file includes:}\\
\indent \indent \indent Materials and Methods \\
\indent \indent \indent Supplementary text\\
\indent \indent \indent Figure~S1, S2, S3, S4, S5\\
\indent \indent \indent Table~S1, S2, S3, S4, S5, S6\\
\indent \indent \indent References ($27$--$65$)}

\clearpage
\section{Data Acquisitions \& Reductions}

The ASAS-SN images were processed by the fully automatic ASAS-SN
pipeline (Shappee et al. in prep.) which uses the ISIS image
subtraction package \cite{alard00}.  Upon careful examination, we
marginally detected ASASSN-15lh in a number of pre-discovery images
from two cameras (5 and 6) on the ASAS-SN southern unit, Cassius.
Fig.~S1 presents the reference images for both cameras constructed
from data obtained between May and December 2014.  These reference
images allow us to remove any host-galaxy light.  To recover a more
robust light curve of the event, we combined individual 90-second
subtracted images that were taken within 7.5 days of each other and
acquired on the same camera (averaging $2-8$ images). 
We find that ASASSN-15lh was first detected in a stack of 6 images acquired between May 4-11, 2015 ($V\sim17.4$ mag).
 The
derived Johnson $V$ magnitudes are reported in Table~S4. In Fig.~S1
we show an example from before and after peak, one from each camera,
demonstrating the quality of our subtractions and detections of
ASASSN-15lh.

Following the discovery of the transient, we obtained a series of 32
{\swift} XRT and UVOT target-of-opportunity (ToO) observations between
2015 June 24 and 2015 September 19. The UVOT observations were
obtained in 6 filters: $V$ (5468~\AA), $B$ (4392~\AA), $U$ (3465~\AA),
$UVW1$ (2600~\AA), $UVM2$ (2246~\AA), and $UVW2$ (1928~\AA)
\cite{poole08}. We used a $5\parcs0$ radius region to extract source counts
and a $\sim40\parcs0$ radius region to extract background counts via the
software task {\sc uvotsource}. We then used the UVOT calibrations
from \cite{poole08} and \cite{breeveld10} to convert the measured
count rates into magnitudes and fluxes. The magnitudes are reported in
the Vega system in Table~S5. We analyzed all available X-ray data obtained by the Swift XRT until 2015 September 18. In this 81 ks exposure we found a 3 sigma upper limit at a level of 3.2$\times 10^{-4}$ counts s$^{-1}$ in the Swift XRT. Using a standard power-law spectrum with a 0.3-10 keV spectral slope $\alpha_{\rm X}$ =1.0 and galactic absorption, this upper limit converts to $1.6\times10^{-14} \,\,{\rm ergs\,s^{-1}\,cm^{-2}}$. The SWIFT multi-band photometry plays an
important in deriving the bolometric luminosity and SED evolutions.
Three months since the first SWIFT observations, the SWIFT colors have
evolved redder and UV fluxes have dropped by $\sim 2 {\,\rm
mag}$. Since phase $t_{\rm test} \sim 50 {\,\rm days}$, the fluxes
from optical and UV have kept approximately constant, entering into a
plateau phase with an increase in UV fluxes near $\sim 80 {\,\rm
days}$ (see also \cite{brownatel,xrayatel}).

We retrieved the reduced (bias subtracted and flat-fielded) LCOGT $BV$
images from the LCOGT image archive. We solved for the astrometry of
the images using astrometry.net \cite{lang10} and ran PSF fitting
photometry with DoPHOT to extract the photometry of ASASSN-15lh. The
magnitude zero points were obtained from photometry of stars in the
field from the AAVSO Photometric All-Sky Survey (APASS;
\cite{henden15}). The magnitudes are reported in Table~S6.

Note that in order to derive the fluxes of the supernova, we subtract
the host galaxy fluxes (derived in the following section) from the
SWIFT and LCOGT for $B$ and $V$ bands. As can be seen from the host SED
discussed in the section below, the host is red and thus its fluxes in
bands bluer than $B$ are negligible as compared to
those of the supernova at the epochs of interest.

The single-slit spectra from the Wide Field Reimaging CCD Camera
(WFCCD) on 2015 June 21 on the du Pont 100-inch Telescope and the Goodman Spectrograph
on the Southern Astrophysical Research Telescope (SOAR) were obtained
at the parallactic angle and reduced with standard routines in the
IRAF {\tt twodspec} and {\tt onedspec} packages.  The reductions
included bias subtraction, flat-fielding, 1-D spectral extraction,
wavelength calibration using an arc lamp, and flux calibration using a
spectroscopic standard usually taken the same night.

The Southern African Large Telescope (SALT) observations were obtained
with the Robert Stobie Spectrograph (RSS), using an atmospheric
dispersion corrector and $1\parcs5$ longslit oriented at a position angle of
127.7$^\circ$ east of north to include a bright reference star on the
slit. The spectrum taken on UT 2015-Jun-30.17 had 2100 sec of total
exposure with the PG0900 grating in two tilt positions to fill chip
gaps and cover the wavelength range 3250-6150 \AA\, at a spectral
resolution $R = \lambda/\Delta\lambda \approx 900$. The UT
2015-07-06.99 spectrum used the PG3000 grating in two tilt positions,
covering the wavelength range 3340-4080 \AA\, with a total exposure of
2400 sec and spectral resolution $R \approx 2600$. These data were
reduced with a custom pipeline that incorporates routines from PyRAF
and PySALT \cite{pysalt}.

We calibrate all spectra by extracting synthetic photometric
magnitudes for each filter in the spectral range and then finding the
best linear fit to the observed magnitudes interpolated to the
spectral epoch. For spectra where only a single filter fit within the
spectral range, the spectrum was scaled by a constant. The calibrated
spectra are shown in Fig.~2 and the digital versions are made
available in the public WISeREP \cite{wiserep} repository.

\section{Host Galaxy Properties}

We searched public data archives and surveys in order to characterize
the candidate host galaxy of ASASSN-15lh, APMUKS(BJ)
B215839.70$-$615403.9 \cite{maddox90}. The host is detected in CRTS
\cite{drake} images and the long term $V$-band light curve does not
show any strong evidence for photometric variability, with a $V$-band
rms of 0.27~mag determined by the photometric uncertainties (see Fig.~S2). 
We found excellent near-infrared $H$ and $K_s$-band
images of the host in the European Southern Observatories (ESO) archive from the VISTA Hemisphere Survey
(VHS; \cite{VHS}). After calibrating the images using 2MASS stars in
the field, we used Sextractor \cite{sextractor} to measure total
aperture magnitudes of the host of $J = 16.04 \pm 0.03$~mag and $K_s =
14.82 \pm 0.04$~mag, which correspond to absolute magnitudes of $M_J
\simeq -24.3$ and $M_{K,s} \simeq -25.5$ for the assumed distance
modulus.

To analyze the SED of the host galaxy, we combined the near-IR fluxes
that we derived from the VHS images with the $griz$ magnitudes of the
host reported from the Dark Energy
Survey (DES) data by \cite{des}. The SED of APMUKS(BJ) B215839.70-615403.9 is shown in
Fig.~S3. We also show a fit using stellar population synthesis (SPS)
models. The fit was obtained using the Fitting and Assessment of
Synthetic Templates (FAST; \cite{kriek09}) code using Bruzual \&
Charlot (2003) SPS models \cite{bc03}, assuming a Salpeter IMF and an
exponentially declining star-formation history ($\rm SFR(t) \propto
e^{-t/\tau}$), and correcting the fluxes for Galactic extinction
only. The best SPS model fit from FAST is excellent, with a reduced
$\chi^2$ of $0.94$, giving a total stellar mass of $\rm
log(M_{\star}/M_{\odot}) = 11.34 \pm 0.1$ ($M_{\star} \simeq 2\times
10^{11}$~M$_{\odot}$), $\rm log(\tau/years) = 9.10^{+0.20}_{-0.08}$,
and $\rm log(age/years) = 9.90^{+0.10}_{-0.04}$. Given the good SPS
fit of the SED, we used this model to derive synthetic magnitudes in
all the Swift/UVOT and LCOGT filters in which we obtained photometry
of ASASSN-15lh, in order to use these fluxes for host-galaxy
subtraction. The resulting synthetic magnitudes of the host are:
$V=18.92$, $B=20.47$, $U=20.69$, $UW1= 21.13$, $UM2 = 21.18$, and
$UW2=21.12$ (Swift/UVOT filters in Vega system), $B=20.37$, $V=18.89$ (Johnson-Kron/Cousins filters in Vega system).

We also analyzed the archival near-IR images of the host with GALFIT
\cite{galfit} in order to fit the morphological structure of the host
galaxy in the 2D images. We find excellent (reduced $\chi^{2}$ of
$\sim$1) Sersic profile fits ($\Sigma(r)\propto r^{-n}$) for the $J$
and $K_s$ bands, without the need of extra components (e.g., a
disk). The results of the fits are: $n=3.41 \pm 0.47$ and $R_e =0\parcs68
\pm 0\parcs03$ in the $J$ band image, $n=4.82\pm 1.16$ and $R_e =0\parcs59 \pm
0.07$ in $K_s$ band image. From these results, the structure of the
host galaxy is consistent with a de~Vaucouleurs spheroidal profile
($n=4$), with an effective radius of $2.4 \pm 0.3 $~kpc.

The high stellar mass, red colors, spheroidal morphology, and low
recent star-formation (obtained from upper limit on [O~II] emission
lines in the supernova spectra) of APMUKS(BJ) B215839.70-615403.9 make
it consistent with an old, massive galaxy. This is in line with the
results of \cite{des}, who found that the SED of the host was that of
a red-sequence galaxy at a photometric redshift of $0.25 \pm 0.02$,
consistent with the spectroscopic redshift measured from the Mg~II
doublet in the supernova spectra. The high stellar mass and low
star-formation rate make the candidate host of ASASSN-15lh distinct
among SLSN-I host galaxies, which tend to be low-metallicity,
low-mass, and high star-formation rate density galaxies (e.g.,
\cite{neillhost,stollhost,chen13,lunnanhost,lunnan15,chen15}. We note that we cannot
rule out the presence of a small dwarf galaxy in projection along the line of sight to 
this massive, compact galaxy with the current images that we have of
the event. Higher resolution data (e.g., with HST) obtained
after the supernova fades will be important for fully characterizing
the host galaxy.

We perform relative astrometry between the LCOGT (June 16, 2015,
$0\parcs47$/pixel, FWHM = 2.3 pixel) and archival DSS image (July 7, 1995,
$1\parcs01$/pixel, FWHM = 5.0 pixel). Using 10 common bright and isolated
stars in the LCOGT field, we establish a 6-parameter coordinate
transformation between the images with IRAF {\tt geomap} package.  The
result of the transformation yields a separation in east (RA)
direction of $0\parcs10\pm 0\parcs13$, and north (Decl.) direction $0\parcs15
\pm 0\parcs10$ between the transformed LCOGT supernova position and the
DSS galaxy centroid position. The supernova was at $V\sim17$ while the
host galaxy has $V\sim 18.9$ so the astrometric separation between the
supernova and galaxy centroid is $0\parcs18 \pm 0\parcs16$.

\section{Blackbody Fits and Total Radiated Energy}

We fit each epoch with photometric data using a simple blackbody model.
For the epochs with Swift data we used no priors on the blackbody temperature
$T_{BB}$.  From these results we designed a simple logarithmic prior for
the epochs with only optical data.  The priors were set to be $20000$~K
on JD 2457191 and $11000$~K on JD 2457241, varying linear between
these dates and constant afterwards.   We then considered two models
for the earlier phases where either the temperature continued to rise
linearly towards earlier times (Table~S2) or was held constant for
the earlier epochs (Table~S3).  The temperature prior was applied to
$\log_{10} T_{BB}$ with an uncertainty of $0.05$~dex.  Parameter
uncertainties were determined using Monte Carlo Markov Chain methods.
The Tables S2 and S3 report the estimated median luminosities, temperatures,
and blackbody radii along with their estimated symmetric 90\% confidence
regions. The total radiated energy $E$ was determined by trapezoidal integration
of the results for the individual epochs.  The uncertainties are dominated
by the choice of the temperature priors, where a shift of $\pm 2000$~K in
the absolute level changes the estimate by $10\%$, and the shift from the
constant temperature prior to the rising temperature prior for early
times increases the estimate by $10\%$.  The span of these systematic
uncertainties leads to our estimate that $E = (1.1 \pm 0.2) \times 10^{52}$~erg.

In Fig.~S4., we show ASASSN-15lh SEDs (filled circles with error bars) at various phases and respective 
best-fit blackbody models (lines). The blackbody models are reasonable descriptions of the SEDs with a well-detected UV peak in the SED that is 
inconsistent with a power-law spectrum.

In Fig.~S5, we compare the derived the bolometric light curves of ASASSN-15lh with those of 
several supernovae of various types. The comparison supernovae include hydrogen-poor 
SLSN iPTF13ajg \cite{iptf13ajg}, PTF12dam \cite{ptf12dam,chen15}, Type Ia supernova
SN 2003du \cite{sn2003du}, broad-line Type Ic (Ic-GRB) supernova (``hypernova'') 
SN 1998bw \cite{sn1998bw}, 
Type IIP supernova SN 1999em \cite{sn1999em} and Type II supernova 
SN 1987A \cite{sn1987a}. Note that the peak bolometric luminosity of ASASSN-15lh is brighter 
than that of the Milky Way\cite{milkyway} by more than one order of magnitude.

\section{Rate}

While ASASSN-15lh is a single event, we can use it to estimate the
rate of such high luminosity SLSNe for comparison to the rate at lower luminosity
found by \cite{rate}. We estimate it by randomly drawing light
curves for the centers of randomly selected $0.01 < z<0.1$ SDSS galaxies observed by the ASAS-SN survey 
for the period 1 June 2014 to 31 May 2015.  The goal here is to roughly sample the 
high latitude sky (e.g. Galactic extinction) and typical weather modulated cadences.
Using $0.01 < z<0.1$ galaxy centers is mildly conservative in that the noise contribution from the
galaxies will be mildly overestimated.  In general, however, since the vast majority
of these galaxies are undetected in ASAS-SN, the use of these galaxies 
differs little from using random positions.  The primary purpose of the procedure
is to supply a random set of observing cadences and their associated noise levels.
Although we survey regions closer to the Galactic plane, we assume the sky 
coverage excludes latitudes $|b|<10^\circ$.
 
To estimate the rates, we randomly select a light curve, a redshift assuming
a constant comoving density in volume $V_{com}$ and the time at which the
transient peaks.  The scrambling of the true SDSS redshifts is unimportant. 
The observed comoving rate and transient time scale are
corrected for time dilation.  Relative to the time at which they peak,
we assume the transients can be modeled as 
\begin{equation}
    M_V(t) = M_{peak} + (t/t_{peak}(1+z))^2 -1.86z 
\end{equation} 
where $t_{peak}$ is the rest frame 
time for the light curve to decay by one magnitude and the last term is a simple 
K-correction matching our estimate for ASASSN-15lh at its observed redshift.  
For ASASSN-15lh, $t_{peak} \simeq 30$ to $40$~days.
Comparing the model fluxes and observed noise at the epochs of the randomly selected light curve, 
we define a detection as an event where the signal-to-noise ratio is $S/N >7.5$ for at 
least two epochs separated by less than one week.  This criterion reproduces the 
observed magnitude distribution of ASAS-SN Type Ia SNe well.  From these statistics
we obtain the average number of days over which we would detect the model transient 
for galaxies randomly distributed out to the distance limit, $\langle t \rangle$,
leading to a rate estimate for $N$ events of $r = N (\langle t \rangle V_{com})^{-1}$.
We tabulated the results for $M_{peak}=-20$, $-21$, $-22$, $-23$ and $-24$
and $t_{peak}=10$, $20$, $30$ and $40$ days. 

For one detected event, the
differential rates for $t_{peak}=40$~days scale with absolute luminosity as 
$r=60$, $15$, $3.8$, $0.96$ and $0.23$~Gpc$^{-3}$~year$^{-1}$ in order
of increasing luminosity.  Because of ASAS-SN's relatively
high cadence, the choice of $t_{peak}$ is unimportant given the
statistical uncertainties provided $t_{peak} \geq 20$~days (using $t_{peak}=30$, $20$ 
or $10$~days instead of $40$~days raises the rate estimates by approximately
3\%, 26\% and 83\%, respectively).  Thus, if we consider ASASSN-15lh as
one event over the magnitude range $-23 < M_{peak} < -24$, the mean rate
is $r\simeq 0.60$~Gpc$^{-3}$~year$^{-1}$ with $0.21 < r < 2.8$ at 90\%
confidence given the Poisson uncertainties for one event.  The lack of events
in the magnitude ranges $-21 < M_{peak} < -22$ and $-22 < M_{peak} < -23$
imply 90\% confidence upper limits on the rates of $r < 22$ and $<5.5$~Gpc$^{-3}$~year$^{-1}$, 
respectively.  This is broadly consistent with (2), who 
found $ r \simeq 32$~Gpc$^{-3}$~year$^{-1}$ ($6 < r < 109$) for SLSNe-I 
concentrated in the (unfiltered) magnitude range $M_{peak} \simeq -21.7 \pm 0.4$~mag,
and the lack of any more luminous examples in \cite{rate} is consistent with our
inference of a steeply falling luminosity function.  Alternately, for
the rate found by (2) we would have expected $1.4$ ($0.3$ to $4.9$) events
with $-21 < M_{\rm peak} < -22$, and our failure to find one is compatible with
Poisson expectations except at the upper end of their rate estimates.  More 
work is required to fully understand the SLSNe luminosity function and to more completely 
quantify the ASAS-SN survey efficiency and completeness.


\section{Alternative Scenarios}

In this section, we examine alternative interpretations of ASASSN-15lh other than a superluminous supernova.

At this low redshift, any significant contribution from gravitational
lensing is completely implausible (see, e.g., the review by Kochanek 2006 \cite{csk06} of lens properties and statistics).  Moreover, the putative
lens (Mg~{\tt \rm II} absorption) and source (broad absorption trough) redshifts
are identical.

That the transient's position is consistent with the center of
the host galaxy raises the possibility that it is some form of
transient created by a supermassive black hole.  Of the
various possibilities, a plausible scenario is that
the transient is a tidal disruption event (TDE) where the
debris from tidally destroying a star is then accreted. Normal 
Active Galactic Nucleus (AGN) activity has been
well-characterized as a stochastic process with well-defined
properties (e.g., \cite{MacLeod14}) that are inconsistent with the observed variability 
 and the spectral slope of the variable
flux is not thermal (e.g., \cite{vandenberk04}).   Only blazars and related ``jetted''
sources show this level of variability, but they also have
featureless, non-thermal, power law spectra (see, e.g.,
\cite{agnbook}). 
The archival CRTS data (Fig. S2) show no variability at the level of $\sim 0.3{\,\rm mag}$ and is thus consistent with no past AGN activities. We cannot rule out the possibility that the host galaxy has a weak AGN with a level of photometric variability below the CRTS detection limits or the existence of weak AGN emission lines that are below our current spectroscopic detection limits. This issue will likely be clarified by further studying the host as ASASSN-15lh fades.

Nuclear outbursts do occur, with the best example being CSS100217
\cite{css10}, a nuclear transient in an already active Seyfert
galaxy. Its nature is debated (SLSN-II or TDE), but it has prominent and changing hydrogen Balmer lines during the transient, which are not seen at all in 
ASASSN-15lh spectra, suggesting that they have different origins.

A fundamental problem for interpreting ASASSN-15lh as a TDE is the
lack of both hydrogen and helium emission lines.  Almost all known
TDEs show varying combinations of strong, broad H$\alpha$/H$\beta$ and
HeII 4686\AA\ emission lines (e.g. \cite{ptftde}).  In the initial
phases, the apparent temperature of ASASSN-15lh combined with its
luminosity implies the generation of large numbers of ionizing photons
($Q \sim 10^{55}$~s$^{-1}$ above 13.6~eV).  A large fraction of these
photons photo-ionize hydrogen and helium and must eventually produce
recombination lines.  The nature of the TDE process makes it very
difficult to avoid having moderate density gas associated with the
disruption that will do the reprocessing.  The balance between
hydrogen and helium can vary both due to the composition of the star
(e.g., the helium star hypothesis for PS1-10jh, \cite{pstde}) and
variations in the effective temperature.  For example,
\cite{Strubbe2015} make very strong arguments that the only way to
avoid having hydrogen emission lines from PS1-10jh is to have no
hydrogen to begin with.  By extension, the only way to avoid having
both hydrogen and helium emission lines from a TDE is to have neither
element, through the disruption of a Wolf-Rayet star
stripped of both elements.   We also note that the UV/optical spectral
energy distributions of TDEs generally are very slowly evolving and
hot (staying $\sim 20000$ to $40000$~K for many months after the peak, see 
\cite{ptftde,14ae,14li}). Their apparent photospheres do not
show the steady cooling and expanding ejecta from explosive events like supernovae  or ASASSN-15lh (see Fig.~3). 

The event called ``Dougie'' \cite{dougie} was an off-nuclear transient
with a featureless spectra.  The temperature evolution from $\sim 13000$~k
to $\sim 6300$~K is typical of supernova and not observed in other 
TDE candidates, both suggesting it was likely a supernova.  
PS1-11af \cite{Chornock14} is regarded as a TDE candidate and 
showed a blue featureless spectra. Its temperature stayed approximately 
constant for three months after the peak, which is typical for  
a TDE but significantly differs from ASASSN-15lh. For both ``Dougie''
and PS1-11af, there are no clear spectral features that resemble  
those seen in ASASSN-15lh.

Finally, the very massive, compact, and old host of ASASSN-15lh would likely imply a large central supermassive black hole mass \cite{haring}. Since TDEs strongly favor relatively low-mass supermassive black holes, this adds to the argument against the TDE interpretation of ASASSN-15lh.

\section*{Further Acknowledgement}
The authors thank PI Neil Gehrels and the Swift ToO team for promptly approving and executing our observations. 
This research has made use of the XRT Data Analysis Software (XRTDAS) developed under the responsibility of the 
ASI Science Data Center (ASDC), Italy. At Penn State the NASA Swift program is support through contract NAS5- 00136. 
We thank LCOGT and its staff for their continued support of ASAS-SN.
This work makes use of observations from the LCOGT network.
This research uses data obtained through the Telescope Access Program (TAP), which has been funded by the ``the Strategic Priority Research Program-The Emergence of Cosmological Structures'' of the Chinese Academy of Sciences (Grant No. XDB09000000) and the Special Fund for Astronomy from the Ministry of Finance.
We thank \'{E}ric Depagne, Marissa Kotze, Paul Kotze, and Brent
Miszalski (SALT staff).
This paper includes data gathered with the 2.5m du Pont Telescope
and the 6.5m Magellan Clay Telescope located at Las Campanas
Observatory, Chile. Based in part on data products from observations made with ESO Telescopes at the Paranal Observatories under ESO program ID 179.A-2010. We also acknowledge CN2015A-85. The LANL LDRD program is funded by US Department of Energy. This research was made possible through the use of the AAVSO Photometric All-Sky Survey (APASS), funded by the Robert Martin Ayers Sciences Fund. This research has made use of the NASA/IPAC Extragalactic Database (NED), which is operated by the Jet Propulsion Laboratory, California Institute of Technology, under contract with the National Aeronautics and Space Administration.

\clearpage
\includegraphics*[width=1.0\textwidth]{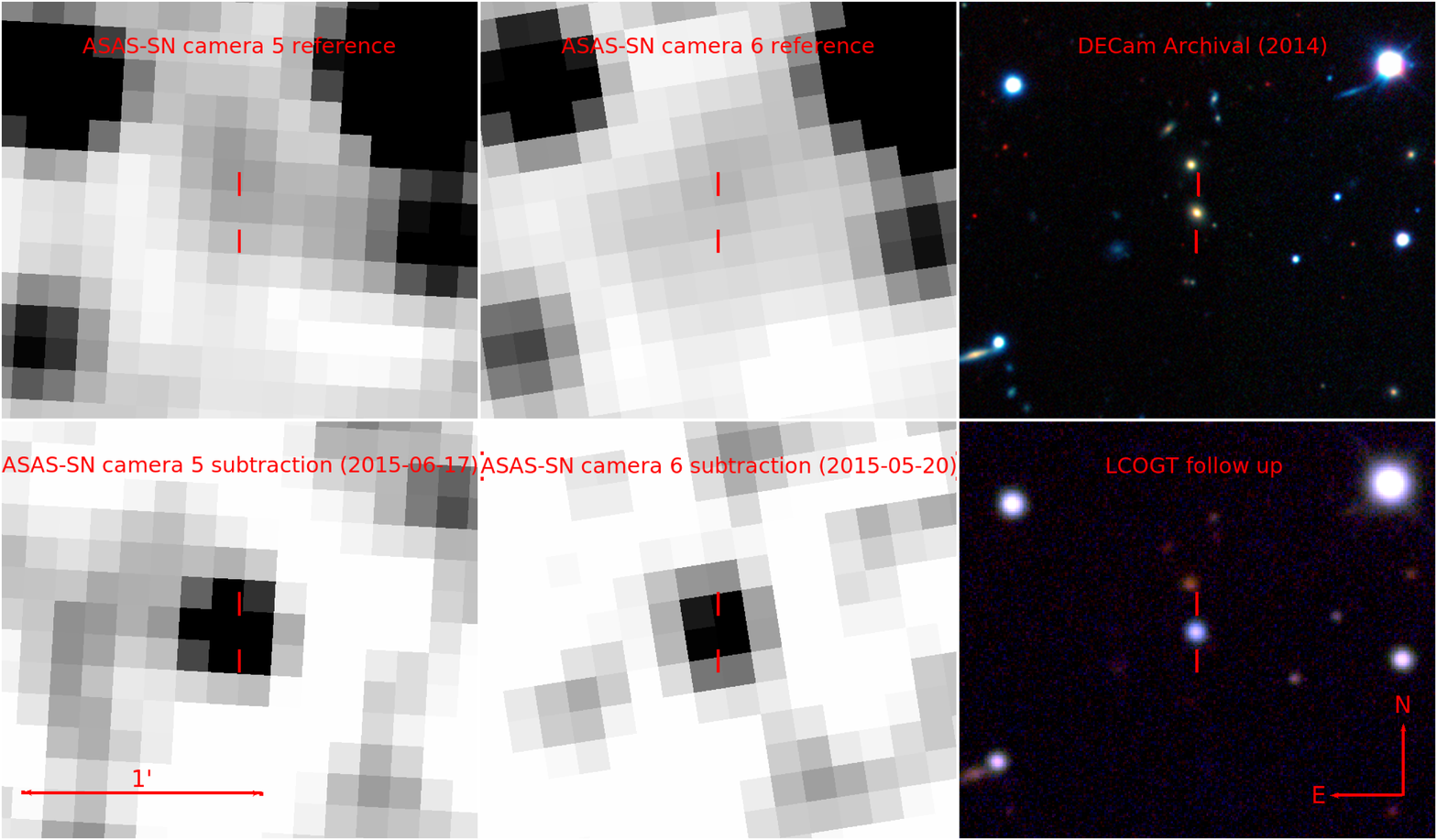}
\noindent{\bf Figure~S1.}
{Comparisons between the ASAS-SN reference image from camera 5 (top-left), 6 (top-middle), the ASAS-SN stacked subtraction image from camera 5 (bottom-left), 6 (bottom-middle) post-maximum, the archival DECam false-color image (\cite{des};{top-right}), and stacked false-color follow-up image from the LCOGT 1-m network ({bottom-right}).  All images are on the same angular scale and the position of ASASSN-15lh is marked.}
\clearpage

\includegraphics*[width=1.0\textwidth]{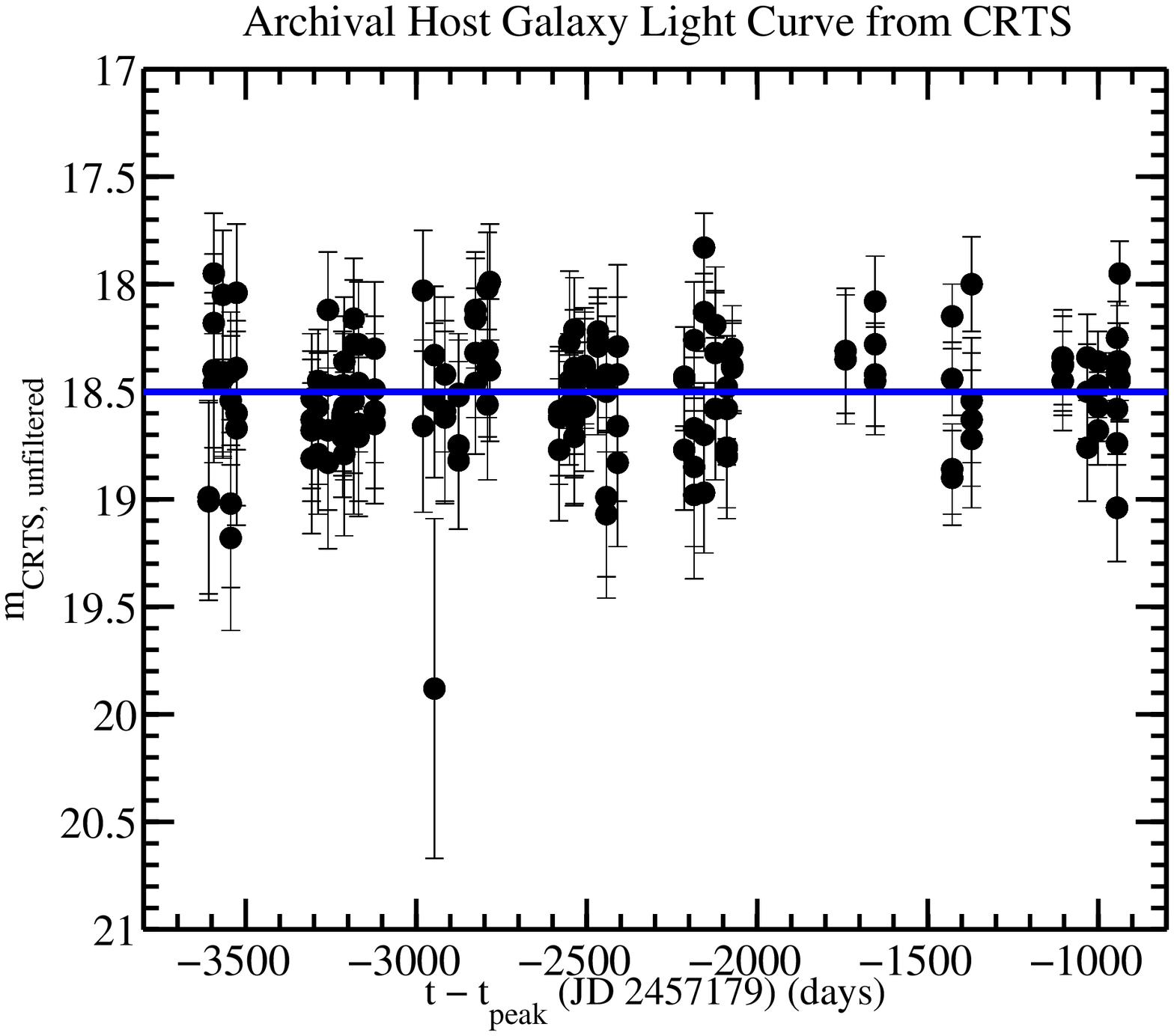}
\noindent{\bf Figure~S2.}  {Pre-explosion archival un-filtered
photometry of APMUKS(BJ) B215839.70-615403.9 (host galaxy of
ASASSN-15lh) from the CRTS survey. The axis shows the time of
observation with respect to the peak date of ASASSN-15lh.  The mean
magnitude is $18.5$ (blue line).  Over $\sim 7.3$\,years of
monitoring, the host is consistent with no photometric variability at
the level of $0.3{\,\rm mag}$.}
\clearpage

\clearpage
\includegraphics*[width=1.0\textwidth]{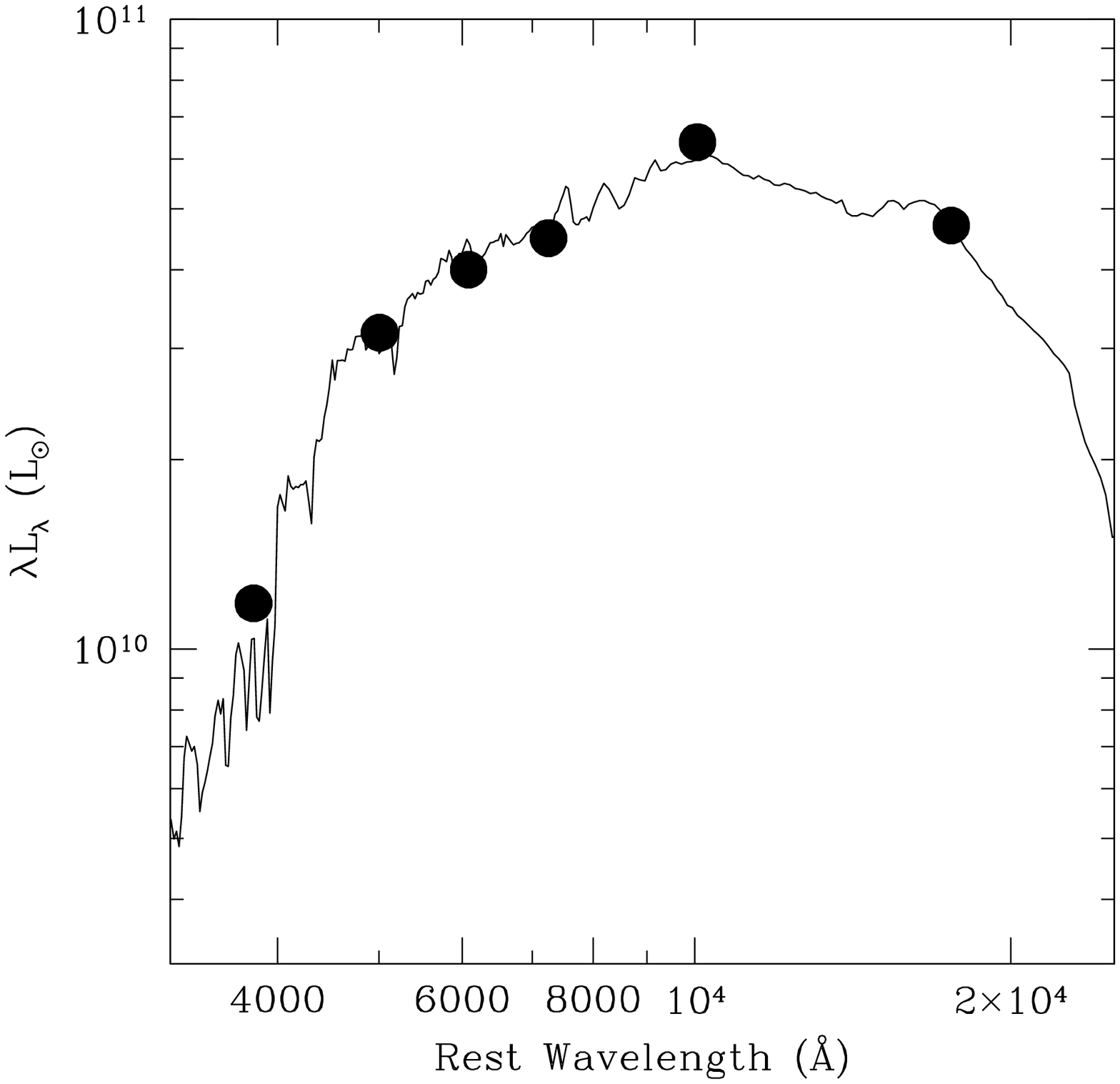}
\noindent{\bf Figure~S3.}  {SED of APMUKS(BJ) B215839.70-615403.9
(host galaxy of ASASSN-15lh) from pre-explosion archival data is shown
in solid dots.  The best-fit galaxy spectrum is shown as a thin solid
line. See Supplementary text for details.}
\clearpage

\clearpage
\includegraphics*[width=1.0\textwidth]{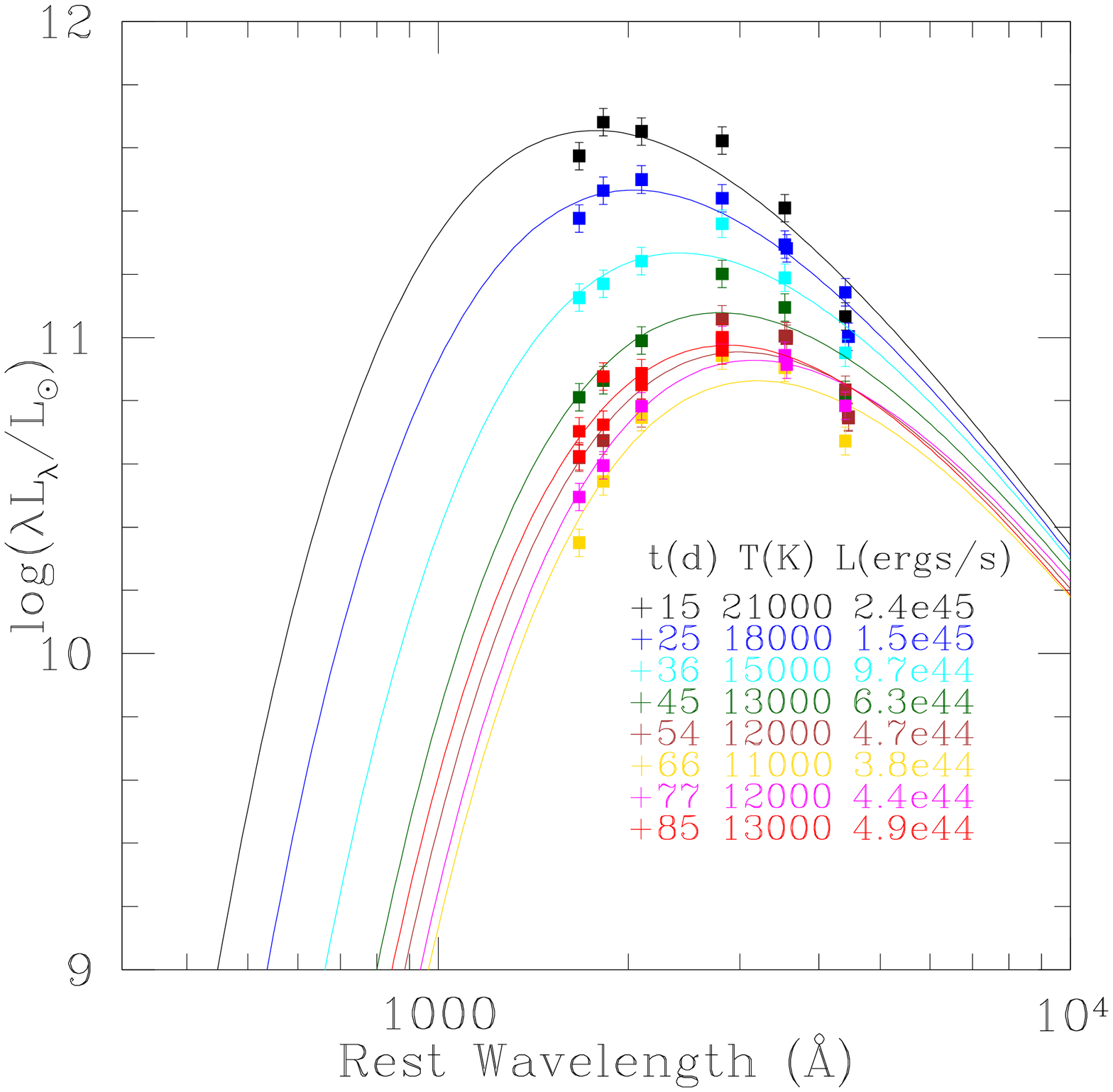}
\noindent{\bf Figure~S4.}  {The evolution of the SED of ASASSN-15lh. Filled circles with error bars show the observed 
SED and the best-fit blackbody models are shown in solid lines.
The best-fit temperatures and luminosities are listed at the 
corresponding  phases.
}
\clearpage

\clearpage
\includegraphics*[width=1.0\textwidth]{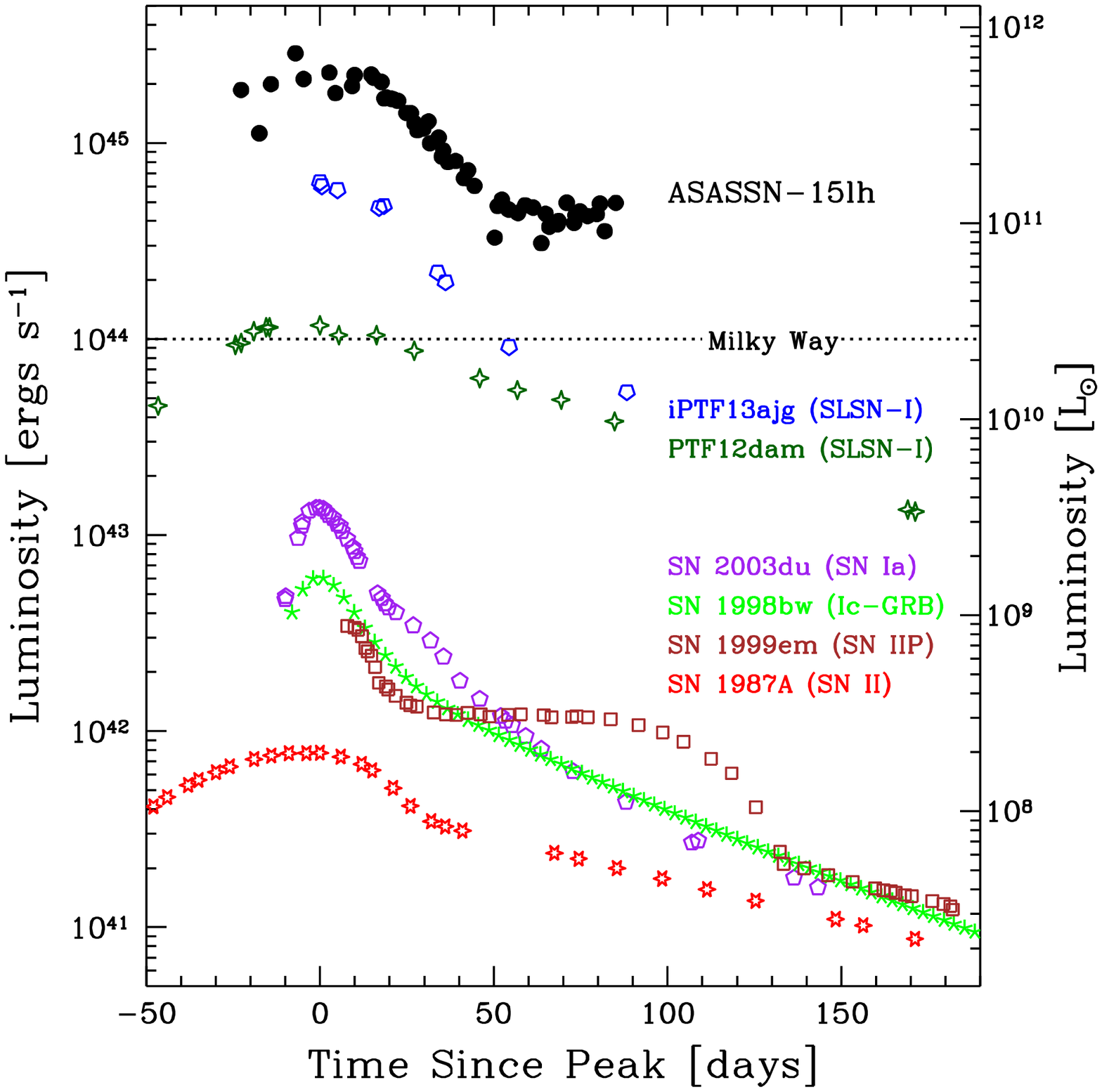}
\noindent{\bf Figure~S5.}  {Bolometric light curves of ASASSN-15lh and other 
supernovae for comparison. The bolometric light curves of hydrogen-poor supernovae iPTFajg, PTF12dam, 
Type Ia supernova SN 2003du, Type Ic-GRB (``hypernova'') SN 1998bw, Type IIP supernova SN 1999em 
and Type II supernova SN 1987A are plotted. The bolometric luminosity of the Milky Way galaxy is 
shown as a dashed line.}
\clearpage

\begin{table}[ht]
\caption*{Table~S1: The Five Most Luminous SNe} 
\centering 
\begin{tabular}{c c c c} 
\hline\hline 
Name & $L_{\rm peak} \,\,{\rm (ergs\,s^{-1})}$& Type  & Ref. \\ [0.5ex] 
\hline 
ASASSN-15lh & $2.2\times10^{45}$ &SLSN-I & This Work\\  
CSS100217&$1.3\times10^{45}$&SLSN-II?&\cite{css10}\\ 
CSS121015&$8.5\times10^{44}$&SLSN-II&\cite{css12}\\
SN 2008es&$6.3\times10^{44}$&SLSN-II&\cite{sn2008es}\\
iPTF13ajg&$6.3\times10^{44}$&SLSN-I&\cite{iptf13ajg}\\ 
\hline 
\end{tabular} 
\label{table:nonlin} 
\end{table} 
We list the peak luminosities and types of the five most luminous SNe discovered to date. ASASSN-15lh 
is $\gtrsim 2$ times more luminous 
than any other SN. Note that CSS100217 is found 
at the core of an AGN, and whether it is a TDE, AGN transient, 
or an SLSN-II is debated (see Supplementary text). 

\begin{table}[ht]
\caption*{Table~S2: Blackbody Temperature $T_{\rm BB}$, Radius 
$R_{\rm BB}$
and bolometric luminosity L derived from SED fitting with  
a linear temperature prior at early time.} 
\centering 
\resizebox{\textwidth}{!}{%
\begin{tabular}{c c c c c c c c c c c} 
\hline\hline 
\mjd     & $N_{\rm obs}$ & $\log L(L_\odot)$ & $\log L_{-}$ & $\log L_{+}$ & $\log T (K)$ & $\log T_{-}$ & $\log T_{+}$ & $\log R (cm)$ & $\log R_{-}$ & $\log R_{+}$ \\
\hline 
57150.5 & 1 & 11.69 & 11.49 & 11.89 & 4.44 & 4.36 & 4.52 & 15.34 & 15.26 & 15.42 \\ 
57157.0 & 1 & 11.46 & 11.28 & 11.67 & 4.41 & 4.33 & 4.50 & 15.28 & 15.20 & 15.36 \\ 
57161.1 & 1 & 11.72 & 11.53 & 11.91 & 4.40 & 4.32 & 4.49 & 15.42 & 15.34 & 15.50 \\ 
57169.8 & 1 & 11.87 & 11.69 & 12.07 & 4.38 & 4.30 & 4.46 & 15.55 & 15.46 & 15.63 \\ 
57172.7 & 1 & 11.74 & 11.56 & 11.94 & 4.37 & 4.28 & 4.45 & 15.51 & 15.42 & 15.59 \\ 
57181.8 & 1 & 11.78 & 11.60 & 11.97 & 4.33 & 4.25 & 4.42 & 15.59 & 15.51 & 15.68 \\ 
57184.0 & 1 & 11.67 & 11.50 & 11.86 & 4.32 & 4.24 & 4.41 & 15.56 & 15.47 & 15.65 \\ 
57189.9 & 2 & 11.71 & 11.55 & 11.88 & 4.30 & 4.22 & 4.38 & 15.62 & 15.53 & 15.71 \\ 
57190.8 & 4 & 11.76 & 11.61 & 11.92 & 4.35 & 4.28 & 4.42 & 15.55 & 15.48 & 15.63 \\ 
57196.7 & 4 & 11.76 & 11.63 & 11.91 & 4.31 & 4.24 & 4.38 & 15.63 & 15.55 & 15.72 \\ 
57197.5 & 6 & 11.75 & 11.71 & 11.79 & 4.29 & 4.26 & 4.33 & 15.66 & 15.60 & 15.71 \\ 
57200.3 & 7 & 11.73 & 11.69 & 11.78 & 4.32 & 4.29 & 4.35 & 15.60 & 15.55 & 15.65 \\ 
57201.3 & 3 & 11.64 & 11.51 & 11.80 & 4.31 & 4.24 & 4.38 & 15.57 & 15.49 & 15.65 \\ 
57202.2 & 6 & 11.65 & 11.62 & 11.68 & 4.23 & 4.20 & 4.26 & 15.73 & 15.67 & 15.79 \\ 
57203.6 & 2 & 11.64 & 11.51 & 11.79 & 4.28 & 4.20 & 4.35 & 15.64 & 15.55 & 15.73 \\ 
57204.7 & 2 & 11.64 & 11.51 & 11.78 & 4.27 & 4.20 & 4.35 & 15.64 & 15.55 & 15.74 \\ 
57206.2 & 8 & 11.63 & 11.60 & 11.66 & 4.26 & 4.23 & 4.29 & 15.67 & 15.62 & 15.72 \\ 
57209.1 & 8 & 11.57 & 11.54 & 11.60 & 4.23 & 4.21 & 4.26 & 15.69 & 15.64 & 15.74 \\ 
57210.7 & 1 & 11.57 & 11.45 & 11.71 & 4.22 & 4.13 & 4.30 & 15.72 & 15.61 & 15.85 \\ 
57212.0 & 6 & 11.52 & 11.49 & 11.55 & 4.20 & 4.17 & 4.22 & 15.73 & 15.68 & 15.79 \\ 
57213.0 & 2 & 11.48 & 11.37 & 11.61 & 4.22 & 4.15 & 4.30 & 15.66 & 15.57 & 15.76 \\ 
57215.2 & 6 & 11.49 & 11.46 & 11.52 & 4.20 & 4.17 & 4.23 & 15.71 & 15.65 & 15.77 \\ 
57217.0 & 2 & 11.53 & 11.46 & 11.60 & 4.13 & 4.09 & 4.17 & 15.88 & 15.78 & 15.98 \\ 
57217.5 & 3 & 11.41 & 11.37 & 11.46 & 4.20 & 4.17 & 4.24 & 15.67 & 15.61 & 15.72 \\ 
57220.6 & 4 & 11.44 & 11.39 & 11.50 & 4.13 & 4.10 & 4.16 & 15.83 & 15.74 & 15.92 \\ 
57221.7 & 2 & 11.35 & 11.24 & 11.47 & 4.19 & 4.12 & 4.27 & 15.65 & 15.56 & 15.76 \\ 
57222.2 & 6 & 11.38 & 11.35 & 11.41 & 4.17 & 4.14 & 4.19 & 15.72 & 15.67 & 15.78 \\ 
57223.9 & 7 & 11.32 & 11.29 & 11.35 & 4.17 & 4.14 & 4.19 & 15.70 & 15.65 & 15.74 \\ 
57226.6 & 6 & 11.32 & 11.29 & 11.36 & 4.11 & 4.09 & 4.14 & 15.80 & 15.75 & 15.86 \\ 
57229.6 & 7 & 11.24 & 11.21 & 11.27 & 4.12 & 4.10 & 4.15 & 15.74 & 15.69 & 15.79 \\ 
57231.0 & 1 & 11.28 & 11.19 & 11.37 & 4.11 & 4.03 & 4.19 & 15.79 & 15.66 & 15.94 \\ 
57233.3 & 6 & 11.20 & 11.16 & 11.23 & 4.10 & 4.08 & 4.13 & 15.76 & 15.71 & 15.81 \\ 
57240.5 & 1 & 10.93 & 10.84 & 11.04 & 4.04 & 3.96 & 4.13 & 15.74 & 15.62 & 15.88 \\ 
57241.5 & 7 & 11.09 & 11.06 & 11.12 & 4.09 & 4.07 & 4.11 & 15.73 & 15.69 & 15.78 \\ 
57243.0 & 2 & 11.13 & 11.05 & 11.22 & 4.10 & 4.03 & 4.17 & 15.74 & 15.63 & 15.85 \\ 
57245.4 & 10 & 11.08 & 11.05 & 11.10 & 4.07 & 4.06 & 4.09 & 15.76 & 15.72 & 15.80 \\ 
57248.7 & 7 & 11.06 & 11.03 & 11.09 & 4.07 & 4.05 & 4.09 & 15.76 & 15.72 & 15.81 \\ 
57251.2 & 8 & 11.10 & 11.07 & 11.13 & 4.06 & 4.04 & 4.08 & 15.80 & 15.76 & 15.84 \\ 
57254.1 & 7 & 11.09 & 11.05 & 11.12 & 4.06 & 4.05 & 4.08 & 15.79 & 15.74 & 15.83 \\ 
57257.0 & 1 & 10.90 & 10.81 & 11.01 & 4.04 & 3.96 & 4.12 & 15.74 & 15.62 & 15.88 \\ 
57258.5 & 2 & 11.05 & 10.98 & 11.15 & 4.11 & 4.04 & 4.18 & 15.68 & 15.58 & 15.80 \\ 
57259.8 & 6 & 10.99 & 10.95 & 11.03 & 4.05 & 4.03 & 4.07 & 15.77 & 15.72 & 15.82 \\ 
\hline 
\end{tabular}} 
\end{table} 
\clearpage
\begin{table}[ht]
\caption*{Table~S2 cont.} 
\centering 
\resizebox{\textwidth}{!}{%
\begin{tabular}{c c c c c c c c c c c} 
\hline\hline 
\mjd     & $N_{\rm obs}$ & $\log L(L_\odot)$ & $\log L_{-}$ & $\log L_{+}$ & $\log T (K)$ & $\log T_{-}$ & $\log T_{+}$ & $\log R (cm)$ & $\log R_{-}$ & $\log R_{+}$ \\
\hline 
57261.6 & 2 & 11.01 & 10.93 & 11.10 & 4.11 & 4.04 & 4.18 & 15.65 & 15.55 & 15.77 \\ 
57262.8 & 2 & 11.00 & 10.93 & 11.08 & 4.07 & 4.00 & 4.14 & 15.72 & 15.61 & 15.85 \\ 
57263.1 & 8 & 11.02 & 10.99 & 11.05 & 4.05 & 4.03 & 4.07 & 15.78 & 15.74 & 15.82 \\ 
57266.0 & 4 & 11.11 & 11.05 & 11.17 & 3.98 & 3.96 & 4.00 & 15.96 & 15.89 & 16.02 \\ 
57268.6 & 2 & 11.01 & 10.94 & 11.10 & 4.10 & 4.03 & 4.17 & 15.67 & 15.56 & 15.78 \\ 
57269.1 & 6 & 11.05 & 11.01 & 11.09 & 4.03 & 4.01 & 4.05 & 15.84 & 15.79 & 15.89 \\ 
57270.7 & 1 & 11.07 & 10.99 & 11.14 & 4.04 & 3.96 & 4.12 & 15.81 & 15.67 & 15.98 \\ 
57273.4 & 7 & 11.04 & 11.01 & 11.08 & 4.05 & 4.04 & 4.07 & 15.78 & 15.74 & 15.83 \\ 
57276.7 & 1 & 11.05 & 10.98 & 11.13 & 4.04 & 3.96 & 4.12 & 15.81 & 15.66 & 15.98 \\ 
57277.8 & 6 & 11.11 & 11.07 & 11.14 & 4.08 & 4.06 & 4.10 & 15.77 & 15.71 & 15.82 \\ 
57279.5 & 2 & 10.97 & 10.89 & 11.05 & 4.10 & 4.03 & 4.17 & 15.66 & 15.55 & 15.77 \\ 
57283.4 & 8 & 11.11 & 11.06 & 11.16 & 4.08 & 4.06 & 4.11 & 15.76 & 15.68 & 15.83 \\ 
\hline 
\end{tabular}} 
\label{table:BBfit} 
\end{table} 

\clearpage
\begin{table}[ht]
\caption*{Table~S3: Blackbody Temperature $T_{\rm BB}$, Radius 
$R_{\rm BB}$
and bolometric luminosity L derived from SED fitting with  
a flat temperature prior at early time.} 
\centering 
\resizebox{\textwidth}{!}{%
\begin{tabular}{c c c c c c c c c c c} 
\hline\hline 
\mjd     & $N_{\rm obs}$ & $\log L(L_\odot)$ & $\log L_{-}$ & $\log L_{+}$ & $\log T (K)$ & $\log T_{-}$ & $\log T_{+}$ & $\log R (cm)$ & $\log R_{-}$ & $\log R_{+}$ \\
\hline 
57150.5 & 1 & 11.39 & 11.22 & 11.58 & 4.30 & 4.22 & 4.38 & 15.47 & 15.38 & 15.56 \\ 
57157.0 & 1 & 11.22 & 11.06 & 11.40 & 4.30 & 4.22 & 4.38 & 15.38 & 15.30 & 15.48 \\ 
57161.1 & 1 & 11.49 & 11.33 & 11.67 & 4.30 & 4.22 & 4.38 & 15.52 & 15.43 & 15.61 \\ 
57169.8 & 1 & 11.71 & 11.54 & 11.89 & 4.30 & 4.22 & 4.38 & 15.62 & 15.54 & 15.72 \\ 
57172.7 & 1 & 11.60 & 11.43 & 11.78 & 4.30 & 4.22 & 4.38 & 15.57 & 15.48 & 15.67 \\ 
57181.8 & 1 & 11.71 & 11.53 & 11.90 & 4.30 & 4.22 & 4.38 & 15.62 & 15.54 & 15.72 \\ 
57184.0 & 1 & 11.62 & 11.45 & 11.80 & 4.30 & 4.21 & 4.38 & 15.58 & 15.49 & 15.68 \\ 
57189.9 & 2 & 11.70 & 11.54 & 11.87 & 4.30 & 4.22 & 4.38 & 15.62 & 15.54 & 15.71 \\ 
57190.8 & 4 & 11.76 & 11.61 & 11.92 & 4.35 & 4.28 & 4.42 & 15.55 & 15.48 & 15.63 \\ 
57196.7 & 4 & 11.76 & 11.63 & 11.91 & 4.31 & 4.24 & 4.38 & 15.63 & 15.55 & 15.72 \\ 
57197.5 & 6 & 11.75 & 11.71 & 11.79 & 4.29 & 4.26 & 4.33 & 15.66 & 15.60 & 15.71 \\ 
57200.3 & 7 & 11.73 & 11.68 & 11.77 & 4.32 & 4.29 & 4.35 & 15.60 & 15.55 & 15.65 \\ 
57201.3 & 3 & 11.65 & 11.51 & 11.80 & 4.31 & 4.24 & 4.38 & 15.57 & 15.49 & 15.66 \\ 
57202.2 & 6 & 11.65 & 11.62 & 11.68 & 4.23 & 4.20 & 4.26 & 15.73 & 15.67 & 15.79 \\ 
57203.6 & 2 & 11.64 & 11.51 & 11.78 & 4.28 & 4.20 & 4.35 & 15.64 & 15.56 & 15.73 \\ 
57204.7 & 2 & 11.63 & 11.51 & 11.78 & 4.27 & 4.20 & 4.35 & 15.64 & 15.56 & 15.74 \\ 
57206.2 & 8 & 11.63 & 11.60 & 11.66 & 4.26 & 4.23 & 4.29 & 15.67 & 15.62 & 15.72 \\ 
57209.1 & 8 & 11.57 & 11.54 & 11.60 & 4.23 & 4.21 & 4.26 & 15.69 & 15.64 & 15.74 \\ 
57210.7 & 1 & 11.57 & 11.45 & 11.71 & 4.22 & 4.14 & 4.30 & 15.72 & 15.61 & 15.84 \\ 
57212.0 & 6 & 11.52 & 11.49 & 11.55 & 4.20 & 4.17 & 4.22 & 15.73 & 15.68 & 15.79 \\ 
57213.0 & 2 & 11.48 & 11.37 & 11.62 & 4.23 & 4.15 & 4.30 & 15.66 & 15.57 & 15.76 \\ 
57215.2 & 6 & 11.49 & 11.46 & 11.52 & 4.20 & 4.18 & 4.23 & 15.71 & 15.65 & 15.77 \\ 
57217.0 & 2 & 11.53 & 11.46 & 11.59 & 4.13 & 4.09 & 4.17 & 15.88 & 15.78 & 15.97 \\ 
57217.5 & 3 & 11.41 & 11.37 & 11.46 & 4.21 & 4.17 & 4.24 & 15.67 & 15.61 & 15.72 \\ 
57220.6 & 4 & 11.44 & 11.38 & 11.50 & 4.13 & 4.10 & 4.17 & 15.83 & 15.74 & 15.92 \\ 
57221.7 & 2 & 11.35 & 11.24 & 11.47 & 4.20 & 4.12 & 4.27 & 15.65 & 15.56 & 15.76 \\ 
57222.2 & 6 & 11.38 & 11.35 & 11.41 & 4.17 & 4.14 & 4.19 & 15.73 & 15.67 & 15.78 \\ 
57223.9 & 7 & 11.32 & 11.29 & 11.35 & 4.17 & 4.14 & 4.19 & 15.70 & 15.65 & 15.74 \\ 
57226.6 & 6 & 11.32 & 11.29 & 11.36 & 4.11 & 4.09 & 4.14 & 15.80 & 15.75 & 15.86 \\ 
57229.6 & 7 & 11.24 & 11.21 & 11.27 & 4.12 & 4.10 & 4.15 & 15.74 & 15.69 & 15.79 \\ 
57231.0 & 1 & 11.28 & 11.19 & 11.38 & 4.11 & 4.03 & 4.19 & 15.79 & 15.66 & 15.94 \\ 
57233.3 & 6 & 11.20 & 11.16 & 11.23 & 4.10 & 4.08 & 4.13 & 15.76 & 15.71 & 15.81 \\ 
57240.5 & 1 & 10.93 & 10.84 & 11.05 & 4.05 & 3.96 & 4.13 & 15.74 & 15.62 & 15.88 \\ 
57241.5 & 7 & 11.09 & 11.06 & 11.12 & 4.09 & 4.07 & 4.11 & 15.73 & 15.69 & 15.78 \\ 
57243.0 & 2 & 11.12 & 11.05 & 11.21 & 4.10 & 4.03 & 4.17 & 15.74 & 15.63 & 15.85 \\ 
57245.4 & 10 & 11.08 & 11.05 & 11.10 & 4.07 & 4.06 & 4.09 & 15.76 & 15.72 & 15.80 \\ 
57248.7 & 7 & 11.06 & 11.03 & 11.09 & 4.07 & 4.05 & 4.09 & 15.76 & 15.72 & 15.81 \\ 
57251.2 & 8 & 11.10 & 11.07 & 11.13 & 4.06 & 4.04 & 4.08 & 15.80 & 15.76 & 15.84 \\ 
57254.1 & 7 & 11.09 & 11.05 & 11.12 & 4.06 & 4.05 & 4.08 & 15.79 & 15.74 & 15.83 \\ 
57257.0 & 1 & 10.90 & 10.81 & 11.02 & 4.04 & 3.96 & 4.12 & 15.74 & 15.62 & 15.88 \\ 
57258.5 & 2 & 11.06 & 10.98 & 11.15 & 4.11 & 4.04 & 4.18 & 15.68 & 15.58 & 15.79 \\ 
57259.8 & 6 & 10.99 & 10.95 & 11.03 & 4.05 & 4.03 & 4.07 & 15.77 & 15.72 & 15.82 \\
\hline  
\end{tabular}} 
\end{table} 
\clearpage
\begin{table}[ht]
\caption*{Table~S3 cont.} 
\centering 
\resizebox{\textwidth}{!}{%
\begin{tabular}{c c c c c c c c c c c} 
\hline\hline 
\mjd     & $N_{\rm obs}$ & $\log L(L_\odot)$ & $\log L_{-}$ & $\log L_{+}$ & $\log T (K)$ & $\log T_{-}$ & $\log T_{+}$ & $\log R (cm)$ & $\log R_{-}$ & $\log R_{+}$ \\
\hline 
57261.6 & 2 & 11.01 & 10.93 & 11.10 & 4.11 & 4.04 & 4.18 & 15.66 & 15.55 & 15.77 \\ 
57262.8 & 2 & 11.00 & 10.94 & 11.08 & 4.07 & 4.00 & 4.14 & 15.72 & 15.61 & 15.85 \\ 
57263.1 & 8 & 11.02 & 10.99 & 11.05 & 4.05 & 4.03 & 4.07 & 15.78 & 15.74 & 15.82 \\ 
57266.0 & 4 & 11.11 & 11.05 & 11.17 & 3.98 & 3.96 & 4.00 & 15.96 & 15.89 & 16.03 \\ 
57268.6 & 2 & 11.01 & 10.93 & 11.10 & 4.10 & 4.03 & 4.17 & 15.67 & 15.56 & 15.79 \\ 
57269.1 & 6 & 11.05 & 11.01 & 11.09 & 4.03 & 4.01 & 4.05 & 15.84 & 15.79 & 15.89 \\ 
57270.7 & 1 & 11.07 & 10.99 & 11.14 & 4.04 & 3.96 & 4.12 & 15.81 & 15.67 & 15.99 \\ 
57273.4 & 7 & 11.04 & 11.01 & 11.08 & 4.05 & 4.04 & 4.07 & 15.78 & 15.73 & 15.83 \\ 
57276.7 & 1 & 11.05 & 10.98 & 11.13 & 4.04 & 3.96 & 4.12 & 15.81 & 15.66 & 15.97 \\ 
57277.8 & 6 & 11.11 & 11.07 & 11.14 & 4.08 & 4.06 & 4.10 & 15.77 & 15.71 & 15.82 \\ 
57279.5 & 2 & 10.97 & 10.89 & 11.05 & 4.10 & 4.03 & 4.17 & 15.66 & 15.55 & 15.77 \\ 
57283.4 & 8 & 11.11 & 11.06 & 11.16 & 4.08 & 4.06 & 4.11 & 15.76 & 15.68 & 15.83 \\ 
\hline 
\end{tabular}} 
\label{table:BBfit} 
\end{table}

\begin{table}[ht]
\caption*{Table~S4: ASAS-SN $V$-band photometry with image subtraction} 
\centering 
\begin{tabular}{c c c c c c c c c c c c c} 
\hline\hline 
 \mjd  &   $V$ &   $e_{V}$\\
\hline 
57110.69520& $>17.64$&  -\\ 
57118.39377& $>17.52$&  -\\ 
57124.75369& $>17.62$&  -\\ 
57127.05225& $>17.82$&  -\\ 
57136.35575& $>17.79$&  -\\ 
57136.88918& $>18.30$&  -\\ 
57150.03689& 17.39&  0.23\\ 
57156.55688& 17.64&  0.19\\ 
57160.65807& 17.22&  0.14\\ 
57169.34084& 16.83&  0.14\\ 
57172.30403& 17.03&  0.18\\ 
57181.35200& 16.82&  0.13\\ 
57183.50502& 16.99&  0.13\\ 
57190.21255& 17.24&  0.16\\ 
57190.29006& 16.94&  0.10\\ 
57200.23218& 17.38&  0.19\\ 
57200.71115& 17.28&  0.24\\ 
\hline 
\end{tabular} 
\label{table:asassn} 
\end{table}

\begin{table}[ht]
\caption*{Table~S5: {\swift} UVOT Photometry} 
{\swift} UVOT Aperture photometry (Johnson-Kron/Cousins) with no 
host subtraction.
\newline
\newline
\centering 
\resizebox{\textwidth}{!}{%
\begin{tabular}{c c c c c c c c c c c c c} 
\hline\hline 
 \mjd  &   $UW2$ &   $e_{UW2}$&   $UM2$ &   $e_{UM2}$&   $UW1$ &   $e_{UW1}$&    $U$  &    $e_{U}$ &   $B$   &    $e_{B}$ &    $V$  &    $e_{V}$\\
\hline 
57197.03&  15.59&   0.05&  15.24&   0.06&  15.29&   0.05&  15.39&   0.04&  16.76&   0.06&  17.07&   0.14\\
57199.76&  15.69&   0.05&  15.32&   0.05&  15.31&   0.05&  15.45&   0.04&  16.75&   0.05&  16.77&   0.10\\
57201.75&  15.97&   0.05&  15.58&   0.06&  15.41&   0.05&  15.52&   0.05&  16.89&   0.06&  16.96&   0.13\\
57205.53&  15.90&   0.05&  15.61&   0.05&  15.56&   0.05&  15.60&   0.05&  16.96&   0.06&  16.92&   0.11\\
57208.60&  16.09&   0.07&  15.78&   0.06&  15.68&   0.07&  15.83&   0.07&  17.02&   0.09&  16.87&   0.17\\
57211.52&  16.34&   0.07&  15.96&   0.06&  15.90&   0.07&  15.75&   0.07&  17.10&   0.09&  17.09&   0.17\\
57214.68&  16.35&   0.04&  16.09&   0.05&  15.94&   0.04&  15.87&   0.04&  17.09&   0.05&  17.20&   0.10\\
57216.50&  16.55&   0.09&  -&  -&  -&  -&  15.90&   0.03&  -&  -&  -&  -\\
57217.38&  16.43&   0.04&  -&  -&  -&  -&  -&  -&  -&  -&  -&  -\\
57219.69&  16.61&   0.10&  -&  -&  16.27&   0.06&  -&  -&  -&  -&  -&  -\\
57220.50&  16.77&   0.09&  -&  -&  -&  -&  16.09&   0.04&  -&  -&  -&  -\\
57221.68&  16.71&   0.06&  16.52&   0.06&  16.31&   0.06&  16.03&   0.05&  17.29&   0.07&  17.31&   0.13\\
57223.48&  16.87&   0.06&  16.69&   0.06&  16.41&   0.06&  16.17&   0.06&  17.36&   0.07&  17.26&   0.13\\
57226.09&  17.14&   0.09&  16.81&   0.10&  16.56&   0.09&  16.28&   0.07&  17.39&   0.09&  17.14&   0.15\\
57229.14&  17.38&   0.11&  16.92&   0.10&  16.67&   0.10&  16.62&   0.12&  17.34&   0.12&  17.53&   0.26\\
57232.80&  17.48&   0.09&  17.27&   0.09&  16.94&   0.09&  16.42&   0.08&  17.50&   0.10&  17.55&   0.20\\
57241.06&  17.84&   0.09&  17.58&   0.08&  17.23&   0.08&  16.67&   0.07&  17.56&   0.08&  17.67&   0.16\\
57244.85&  17.94&   0.10&  17.73&   0.09&  17.50&   0.10&  16.77&   0.08&  17.72&   0.09&  17.52&   0.16\\
57248.37&  18.10&   0.10&  17.77&   0.10&  17.47&   0.10&  16.77&   0.09&  17.60&   0.10&  17.68&   0.20\\
57250.71&  18.07&   0.10&  17.81&   0.10&  17.31&   0.10&  16.83&   0.09&  17.50&   0.09&  17.36&   0.16\\
57253.62&  18.10&   0.13&  17.77&   0.12&  17.37&   0.13&  16.63&   0.10&  17.69&   0.12&  17.39&   0.21\\
57259.27&  18.57&   0.16&  18.04&   0.13&  17.53&   0.13&  17.05&   0.12&  17.95&   0.14&  17.80&   0.27\\
57262.59&  18.34&   0.16&  18.11&   0.14&  17.47&   0.13&  17.03&   0.13&  17.82&   0.14&  17.59&   0.25\\
57265.53&  18.91&   0.33&  -&  -&  17.52&   0.10&  17.00&   0.09&  17.97&   0.11&  -&  -\\
57268.65&  18.45&   0.12&  18.06&   0.10&  17.62&   0.10&  17.02&   0.09&  17.83&   0.10&  17.58&   0.17\\
57272.84&  18.14&   0.10&  17.98&   0.16&  17.53&   0.10&  16.99&   0.09&  17.84&   0.11&  17.64&   0.19\\
57277.29&  17.84&   0.08&  17.67&   0.10&  17.15&   0.08&  16.89&   0.06&  17.74&   0.07&  17.53&   0.11\\
57282.81&  17.75&   0.10&  17.61&   0.13&  17.19&   0.10&  16.91&   0.09&  -&  -&  -&  -\\
57283.07&  17.94&   0.10&  17.24&   0.11&  17.27&   0.10&  17.02&   0.09&  -&  -&  -&  -\\
57284.21&  17.75&   0.10&  17.36&   0.09&  17.18&   0.10&  16.98&   0.10&  17.85&   0.12&  17.95&   0.25\\
\hline 
\end{tabular}}
\label{table:swift} 
\end{table} 

\begin{table}[ht]
\caption*{Table~S6: LCOGT Photometry} 
LCOGT DoPHOT photometry (Johnson-Kron/Cousins) with no 
host subtraction.
\newline
\newline
\centering 
\begin{tabular}{c c c c c c c c c c c c c} 
\hline\hline 
     \mjd &     $B$ &    $e_B$ &    $V$ &   $e_V$\\
     \hline 
57189.444&  -&  -&  16.91&   0.03\\
57190.421&  16.74&   0.07&  16.92&   0.05\\
57196.164&  16.79&   0.04&  16.92&   0.03\\
57200.804&  16.86&   0.06&  17.00&   0.04\\
57203.138&  16.90&   0.06&  17.06&   0.04\\
57204.193&  16.91&   0.06&  17.06&   0.04\\
57206.194&  16.95&   0.06&  17.13&   0.05\\
57208.625&  17.04&   0.05&  17.15&   0.03\\
57210.260&  17.03&   0.06&  -&  -\\
57212.524&  17.15&   0.06&  17.16&   0.04\\
57216.869&  17.18&   0.07&  17.27&   0.07\\
57221.188&  17.33&   0.06&  17.36&   0.04\\
57222.884&  -&  -&  17.43&   0.05\\
57228.724&  17.52&   0.08&  -&  -\\
57230.505&  17.46&   0.09&  -&  -\\
57240.034&  -&  -&  17.65&   0.04\\
57241.011&  -&  -&  17.60&   0.04\\
57242.508&  17.62&   0.06&  17.53&   0.04\\
57244.977&  17.70&   0.04&  17.62&   0.03\\
57247.544&  -&  -&  17.61&   0.04\\
57250.526&  17.78&   0.07&  17.71&   0.04\\
57253.515&  -&  -&  17.72&   0.04\\
57256.599&  -&  -&  17.69&   0.06\\
57258.035&  17.75&   0.09&  17.70&   0.07\\
57261.087&  17.86&   0.07&  17.79&   0.06\\
57262.287&  17.95&   0.11&  17.64&   0.10\\
57262.616&  17.79&   0.07&  17.77&   0.07\\
57268.141&  17.87&   0.05&  17.75&   0.03\\
57270.208&  17.88&   0.05&  -&  -\\
57273.206&  17.89&   0.05&  -&  -\\
57276.279&  17.91&   0.05&  -&  -\\
57279.015&  17.98&   0.08&  17.80&   0.05\\
\hline 
\end{tabular} 
\label{table:lcogt} 
\end{table} 

\clearpage



\end{document}